\documentclass[11pt]{article}
\usepackage{latexsym,amssymb,amstext,amsmath}
\usepackage{slashed}
\usepackage{mathrsfs}
\usepackage{hyperref}
\allowdisplaybreaks
\def\slash#1{\rlap{\hbox{$\mskip 1 mu /$}}#1}      
\def\Slash#1{\rlap{\hbox{$\mskip 3 mu /$}}#1}      
\newcommand{\ft}[2]{{\textstyle\frac{#1}{#2}}}

\newcommand {\cC}{{\cal C}}
\newcommand {\cD}{{\cal D}}
\newcommand {\cE}{{\cal E}}

\newcommand {\cH}{{\cal H}}

\newcommand {\cL}{{\cal L}}

\newcommand {\cO}{{\cal O}}

\newcommand {\cR}{{\cal R}}

\newcommand {\cV}{{\cal V}}

\def\l{\lambda}

\def\q{\theta}

\def\z{\zeta}

\def\L{\Lambda}

\def\ri{{\rm i}}

\newcommand{\pa}{\partial}

\newcommand{\be}{\begin{equation}}
\newcommand{\ee}{\end{equation}}
\newcommand{\bea}{\begin{eqnarray}}
\newcommand{\eea}{\end{eqnarray}}

\newcommand{\ba}{\begin{array}}
\newcommand{\ea}{\end{array}}

\def\double #1{#1{\hbox{\kern-2pt $#1$}}}

\newcommand{\bsubeq}{\begin{subequations}}
\newcommand{\esubeq}{\end{subequations}}

\newcommand{\rd}{\mathrm d}
\newcommand{\HC}{{\mathrm{h.c.}}}
\newcommand{\veps}{\varepsilon}

\newcommand{\eps}{{\epsilon}}

\newcommand{\Tr}{\mathrm{Tr }}

\newcommand{\eol}{\nonumber \\}

\newcommand{\wt}{\breve}

\begin{document}
%
\begin{titlepage}
\begin{flushright} \small
  Nikhef-2014-001 \\  ITP-UU-14/01 
\end{flushright}
\bigskip

\begin{center}
  {\LARGE\bfseries Non-renormalization theorems and\\ 
    N=2 supersymmetric  backgrounds}
  \\[10mm]
\textbf{Daniel Butter$^{a}$, Bernard de Wit$^{a,b}$ and Ivano Lodato$^{a}$}\\[5mm]
\vskip 6mm
$^a${\em Nikhef, Science Park 105, 1098 XG Amsterdam, The
  Netherlands}\\
$^b${\em Institute for Theoretical Physics, Utrecht
  University,} \\
  {\em Leuvenlaan 4, 3584 CE Utrecht, The Netherlands}\\[3mm]
 {\tt dbutter@nikhef.nl}\,,\;  {\tt B.deWit@uu.nl}\,,\; {\tt ilodato@nikhef.nl} 
\end{center}

\vspace{3ex}

\begin{center}
{\bfseries Abstract}
\end{center}
\begin{quotation} \noindent The conditions for fully supersymmetric
  backgrounds of general $N\!=\!2$ locally supersymmetric theories are
  derived based on the off-shell superconformal multiplet
  calculus. This enables the derivation of a non-renormalization
  theorem for a large class of supersymmetric invariants with
  higher-derivative couplings. The theorem implies
  that the invariant and its first order variation must vanish in a fully
  supersymmetric background. The conjectured relation of one
  particular higher-derivative invariant with a specific
  five-dimensional invariant containing the mixed gauge-gravitational
  Chern-Simons term is confirmed.
\end{quotation}

\vfill

\flushleft{\today}
\end{titlepage}

\section{Introduction}
\label{sec:introduction}
\setcounter{equation}{0}
There is increasing interest in locally supersymmetric actions with
higher-derivative couplings, whose rigorous study is possible in the
context of a consistent off-shell formulation.
Such formulations are available when the number of supersymmetries is
less than or equal to eight. An off-shell analysis of partially or
fully supersymmetric backgrounds is then feasible and the results
thereof are relevant for various applications. A first step towards
this was made some time ago in \cite{LopesCardoso:2000qm} in the
context of evaluating the corrections to BPS black hole entropy from a
specific higher-derivative coupling. More recent results concern the
discovery of so-called non-renormalization theorems according to which
certain classes of actions as well as their first derivatives with
respect to fields or coupling constants must vanish in a fully
supersymmetric background \cite{deWit:2010za,Butter:2013lta}. This
implies that those actions will not contribute to BPS black hole
entropy and neither do they contribute to the field equations when
studying supersymmetric field configurations.

In flat space-time the analysis of fully supersymmetric backgrounds is
rather straightforward. In that case the supersymmetry algebra
generically implies that all component fields are space-time
independent, so that all derivative terms in the supersymmetry
transformations can be ignored.  It then follows that all fields that
are in the image of the supercharges must vanish. Therefore only
the lowest-dimensional field, which cannot be generated by applying a
supersymmetry transformation on yet another field, can take a finite,
but constant value. In terms of superfields, this means that full
supersymmetry requires any superfield to be constant, i.e. independent
of both the bosonic and the fermionic coordinates.  In the context
of non-trivial space-times, similar results can be derived as long as
one is dealing with rigid supersymmetry.

The first part of this paper deals with a systematic analysis of the
supersymmetric values that certain supermultiplets can take, but now
in the context of local supersymmetry which is somewhat more
subtle. When considering a large variety of supersymmetric invariants,
we prefer to make use of the (off-shell) superconformal multiplet
calculus, where one encounters an extended set of local gauge
invariances associated with the superconformal algebra. Proper
attention should be paid to all these invariances.  This last aspect
does not form an impediment for analyzing supersymmetric backgrounds
and in fact the presence of the extra conformal (super)symmetries
greatly improves the systematics of the analysis.  But it is important
to appreciate that we are now dealing with {\it local} gauge
invariances which imply a reduction of the physical degrees of
freedom. Therefore it does not make sense to just impose gauge
invariance on a field configuration and it is natural that a gauge
invariant orbit of solutions will remain at the end. In principle this
implies that a fully supersymmetric background is only determined up
to (small) gauge transformations. In practice this means that we will
obtain (conformally) covariant conditions on the field configuration.

This is perhaps the point to briefly introduce the various gauge
invariances belonging to the superconformal group. There are two types
of supersymmetries, called Q- and S-supersymmetry. Furthermore there
are space-time diffeomorphisms, local Lorentz transformations (M),
dilatations (D), special conformal boosts (K), and finally the local
R-symmetry transformations that constitute the group
$\mathrm{SU}(2)\times\mathrm{U}(1)$. In the superconformal setting a
(conformal primary) superfield is characterized by its behaviour under
dilatations and the local R-symmetry.  The behaviour under dilatations
and $\mathrm{U}(1)$ transformations is generally characterized by the
so-called Weyl and chiral weights, $w$ and $c$, respectively.

To explain the strategy that we will follow in this paper for
establishing supersymmetric backgrounds and to further elucidate some
of the conceptual issues, we start in section
\ref{sec:vector-supermultiplet} by discussing a single $N=2$ vector
supermultiplet coupled to a conformal supergravity background (whose
covariant quantities comprise the so-called Weyl multiplet). When
deriving the consequences of supersymmetry for the resulting field
configuration we naturally discover that the conformal supergravity
background itself is also subject to constraints. These constraints are
identical to the ones that apply to the Weyl multiplet without the presence of the vector
multiplet. 

In section \ref{sec:short-mutiplets}, we briefly present
three other short supermultiplets coupled to a conformal supergravity background,
namely the tensor multiplet, the non-linear multiplet, and the
hypermultiplet. These three multiplets are all characterized by the
fact that their lowest-weight scalars transform under the
$\mathrm{SU}(2)$ R-symmetry group. Requiring supersymmetry in the
presence of any of these multiplets turns out to impose a stronger
restriction on the Weyl multiplet than when only vector multiplets are
present. With this additional restriction the allowed field
configurations are equivalent to the ones derived in
\cite{LopesCardoso:2000qm}. 

Having determined the conditions imposed by supersymmetry we turn to a
large class of supersymmetric actions with higher-derivative
couplings. We first concentrate on the kinetic multiplet of the
logarithm of a conformal primary anti-chiral superfield of Weyl weight
$w$, $\mathbb{T} (\ln \bar\Phi_w)$. This multiplet has been
extensively discussed in \cite{Butter:2013lta}. The superfield
$\bar\Phi_w$ is usually not an elementary multiplet but a composite
one, and the kinetic multiplet plays a role in constructing a class of
higher-derivative supersymmetric actions that extend the class studied
in \cite{deWit:2010za} which corresponds to the case of $w=0$. One
such action seems to emerge upon dimensional reduction from the
higher-derivative coupling constructed in five dimensions in
\cite{Hanaki:2006pj}. This was first noted in \cite{Banerjee:2011ts}
but at that time only the $w=0$ version of $\mathbb{T} (\ln
\bar\Phi_w)$ was known. In \cite{Butter:2013lta} the construction of
$\mathbb{T} (\ln\bar\Phi_w)$ was presented for arbitrary values of
$w$, and it was concluded that the actual invariant arising from
dimensional reduction corresponds to the case with $w=1$.  To exhibit
some characteristic features of these couplings one may consider the
purely bosonic case, where the relevant expression that appears in the
action equals
\begin{align}
  \label{eq:box-box-log}
  \Box_\mathrm{c} \Box_\mathrm{c} \ln\phi =&\,
  \big(\mathcal{D}^2\big){}^2\ln\phi - 2\, \mathcal{D}^\mu \big[\big(2\,
  f_{(\mu}{}^a e_{\nu)a} - f\,g_{\mu\nu}\big)
  \mathcal{D}^\nu\ln\phi\big]
	\nonumber \\ & \,
  +  w \big[\mathcal{D}^2 f + 2\, f^2 -2\, (f_\mu{}^a)^2 \big]\,.
\end{align}
The scalar field $\phi$ can be either an elementary or a composite
field, and it scales under local dilatations according to $\phi\to
\exp[w\,\Lambda_\mathrm{D}]\,\phi$, where $w$ denotes the (arbitrary)
scaling weight of the field. The derivatives are standard
gravitational derivatives and $f_\mu{}^a$ is a composite gauge field
associated with special conformal boosts, which, in the simple theory
introduced above with a gravitational background, can be expressed in
terms of the Riemann tensor. In that case one has the identity
\begin{equation}
    \label{eq:w-terms-in-R}
    \mathcal{D}^2 f + 2\, f^2 -2\, (f_\mu{}^a)^2 = 
     \tfrac16\mathcal{D}^2 \mathcal{R}  
    -\tfrac12  \mathcal{R}^{ab} \,\mathcal{R}_{ab} +\tfrac16 \mathcal{R}^2~,
\end{equation}
where $\mathcal{R}_{ab}$ and $\mathcal{R}$ denote the Ricci tensor and
scalar. The crucial property of this expression is that it is
conformally invariant irrespective of the value of the Weyl weight and
furthermore that it can be easily extended to $N=2$ supergravity on
the basis of chiral supermultiplets. Hence this expression defines
a class of actions upon multiplying with any (composite or elementary)
scalar of weight $w=0$.

In section \ref{sec:T-log-chiral} we summarize the salient features of
the chiral multiplet $\mathbb{T} (\ln \bar\Phi_w)$ and derive the
conditions imposed by full supersymmetry.  This then facilitates our
task, undertaken in section \ref{sec:non-renormalization}, to
establish the existence of the non-renormalization theorem of the type
discussed before for this class of couplings. This result thus
establishes an extension of the non-renormalization theorem that was
initially proven for the more restricted class of higher-derivative
couplings with $w=0$ \cite{deWit:2010za}. Some early indications of
this extended non-renormalization theorem were already noted in
\cite{Butter:2013lta}, where some applications were also pointed out. 

In section \ref{sec:5d4d}, we return to the issue of the dimensional
reduction of the supersymmetric $5D$ mixed gauge-gravitational
Chern-Simons invariant given in \cite{Hanaki:2006pj}. The resulting
$4D$ action has two contributions: one is a holomorphic term involving
the square of the Weyl multiplet, and the other involves the new
higher-derivative coupling discussed above. Its existence confirmed
the observation made in a study of $5D$ BPS black holes and black
rings in the context of a Lagrangian with the same $5D$
higher-derivative couplings, that the $5D$ equations of motion do not
reduce to the expected $4D$ equations, thus indicating the presence of
new $4D$ higher-derivative couplings \cite{Castro:2008ne}.  In
\cite{Banerjee:2011ts} these new $4D$ couplings were
identified with those constructed in \cite{deWit:2010za}, which
involve the $w=0$ version of $\mathbb{T} (\ln\bar\Phi_w)$. The more
general class based on $w\not=0$ was considered later in
\cite{Butter:2013lta}, and at that point it was noted that actually
the new higher-derivative coupling should correspond to the case
$w=1$. However, a comprehensive proof of this correspondence was
missing until now, and this is the reason why this topic is addressed
in this last section.

For further definitions and notational details, we refer the reader
to the literature, and in particular to \cite{deWit:2010za, Butter:2013lta}.

\section{Vector supermultiplets in a superconformal background}
\label{sec:vector-supermultiplet}
\setcounter{equation}{0}
In this section we derive the conditions that follow from imposing
full supersymmetry on a field configuration consisting of a single
vector supermultiplet in a conformal supergravity background. We first
focus on the conditions imposed by supersymmetry on the vector
multiplet. This eventually leads to conditions on the Weyl multiplet,
the supermultiplet that characterizes the conformal supergravity
background. The same analysis for the Weyl supermultiplet without any
vector multiplet present turns out to lead to identical
conditions. This situation will change in the case that other
supermultiplets than the vector one are present, as will be shown in
section \ref{sec:short-mutiplets}. There we will deal with the
remaining short supermultiplets, namely the tensor multiplet, the
so-called non-linear multiplet and the hypermultiplet. As it turns
out, in the presence of either one of these multiplets, the Weyl
multiplet is subject to additional restrictions.

The vector multiplet consists of a complex scalar $X$, transforming
with weights $w=1$ and $c=-1$ under local dilatations and chiral
$\mathrm{U}(1)$ transformations, a Majorana spinor doublet decomposed
into chiral and anti-chiral components, $\Omega_i$ and $\Omega^i$,
which are each other's conjugates, an abelian gauge field $W_\mu$ and
a triplet of auxiliary fields $Y^{ij}$. The indices $i,j,\ldots=1,2$
refer to the components of the doublet representation of the
R-symmetry group $\mathrm{SU}(2)$. For further definitions we refer
the reader to, for instance, \cite{deWit:2010za,Butter:2013lta}, where
explicit definitions and further details are given in the same
notation as employed in this paper. Under Q- and S-supersymmetry the
transformation rules of the vector multiplet take the following form:
\begin{align}
  \label{eq:variations-vect-mult}
  \delta X =&\, \bar{\epsilon}^i\Omega_i \,,\nonumber\\
  \delta\Omega_i =&\, 2 \Slash{D} X\epsilon_i
     +\ft12 \varepsilon_{ij}  \hat F_{bc}^-
   \gamma^{bc}\epsilon^j +Y_{ij} \epsilon^j
     +2X\eta_i\,,\nonumber\\
  \delta W_{\mu} = &\, \varepsilon^{ij} \bar{\epsilon}_i
  (\gamma_{\mu} \Omega_j+2\,\psi_{\mu j} X)
  + \varepsilon_{ij}
  \bar{\epsilon}^i (\gamma_{\mu} \Omega^{j} +2\,\psi_\mu{}^j
  \bar X)\,,\nonumber\\
\delta Y_{ij}  = &\, 2\, \bar{\epsilon}_{(i}
  \Slash{D}\Omega_{j)} + 2\, \varepsilon_{ik}
  \varepsilon_{jl}\, \bar{\epsilon}^{(k} \Slash{D}\Omega^{l)} \,.
\end{align}
The derivatives $D_\mu$ are fully covariant with respect to superconformal
transformations and thus contain the various connection fields
associated with the superconformal gauge symmetries. The parameters of
Q- and S-supersymmetry are the chiral spinors $\epsilon^i$ and
$\eta_i$, respectively, and their conjugate (anti-chiral) spinors,
$\epsilon_i$ and $\eta^i$. We should point out that $\hat
F^\pm_{\mu\nu}$ are the (anti-)selfdual components of the modified
field strength tensor associated with the gauge field $W_\mu$,
\begin{equation}
  \label{eq:vector-field-strength}
  \hat F_{\mu\nu} = \partial_\mu W_\nu - \partial_\nu W_\mu  -
  \tfrac14\big[ X \,T_{\mu\nu\,ij}\,\varepsilon^{ij}  +\bar X
  \,T_{\mu\nu}{}^{ij}\,\varepsilon_{ij} \big] \,,
\end{equation}
up to additional contributions quadratic in fermion fields. The fields
$T_{ab\,ij}$ and $T_{ab}{}^{ij}$ are the self-dual and anti-selfdual
covariant tensor fields that belong to the Weyl multiplet. Note that
we will generally suppress terms that are of higher order in the
fermions, because eventually the supersymmetric field configurations
will be presented with all fermion fields set to zero.

Before beginning the actual analysis of supersymmetric field
configurations, let us recall that the superconformal symmetries are
realized as {\it local} gauge invariances, which makes the analysis
conceptually rather different as compared to the rigid case. For
instance, imposing rigid supersymmetry requires the scalar field $X$
to be constant.  In the present context such a result is not
meaningful, because $X$ is subject to local scale and phase
transformations, so that any two non-zero values of the field $X$ will
be gauge equivalent.  A similar comment applies also to the fermions,
where one might expect that the fields $\Omega_i$ will be required to
vanish. But here again one realizes that two different values of
$\Omega_i$ can be gauge equivalent by S-supersymmetry. Obviously a
gauge invariant orbit of solutions must remain, but it is often
convenient to choose a particular representative of the gauge orbit, which
is equivalent to adopting a gauge condition. However, we prefer to
restrict this option to the fermionic symmetries and leave the bosonic
superconformal gauge invariances unaffected to keep the structure of
our results as transparent as possible.

Let us now point out that in certain cases the analysis of
supersymmetric configurations can be more direct, which is an
important result that will be relevant throughout this paper. Rather
than considering a single vector multiplet, let us briefly consider
two such multiplets with fields $(X^1,X^2)$,
$(\Omega_i{}^1,\Omega_i{}^2)$, etcetera. Then we may consider a
(conformal primary) chiral multiplet with the components
\begin{equation}
  \label{eq:ratio-chiral}
  \frac{X^1}{X^2}\;, \quad \frac{X^2\,\Omega_i{}^1 -
    X^1\,\Omega_i{}^2}{(X^2)^{2}}\;,\quad \mbox{etcetera}\,. 
\end{equation}
Now the analysis of full supersymmetry becomes straightforward,
because the first (scalar) component is invariant under dilatations and
$\mathrm{U}(1)$ transformations (it has weights $w=c=0$), whereas the
second fermionic component is invariant under S-supersymmetry.
Therefore it is now straightforward to conclude that the scalar must be a
constant, while the fermionic component must vanish. Continuing this
analysis will show that this multiplet is restricted to a 
constant, or, equivalently, that in the supersymmetric limit the two
multiplets must be proportional to one another. This is an example of a
more generic result: if the lowest-weight (scalar) component of a
multiplet does not transform under dilatations and $\mathrm{U}(1)$
transformations, then the supersymmetry algebra implies that the
lowest-weight fermion into which it transforms must be invariant under
S-supersymmetry. In the supersymmetric limit, this multiplet is then
restricted to a constant. For a general chiral multiplet this result was
proven in \cite{deWit:2010za}.

From the above result it is therefore clear that nothing will be
learned by considering several vector multiplets at once, so we return
to the original problem using to a single vector multiplet. Given the
fact that the local superconformal gauge invariances will naturally
lead to a certain degeneracy, we will define a specific approach based
on two guiding principles. First of all, we insist that the bosonic
superconformal invariances are preserved so that the final result can
be expressed in terms of equations that are manifestly covariant with
respect to all these gauge invariances. Secondly we assume that all
(supercovariant) fermionic quantities will vanish in the bosonic
background. This leaves the bosonic invariance intact. The only
equations that are relevant thus follow from the requirement that the
supersymmetry variations of the (supercovariant) fermionic quantities
should vanish under a particular set of supersymmetry transformations
parametrized by eight independent spinorial parameters $\epsilon^i$
and $\epsilon_i$. The resulting bosonic covariant equations then
characterize all the supersymmetric configurations. As we shall see,
this strategy amounts to choosing a certain representative of the
fermionic gauge orbit. In principle one can still apply the fermionic
gauge transformations, but this will then lead to a different
representative for which the fermion fields do not vanish.

Hence, in order that $X$ is invariant under full supersymmetry one
naturally assumes that $\Omega_i=0$. To
ensure that the transformation of the fermions will vanish as well, one
requires that a linear combination of Q- and S-supersymmetry will
vanish on the spinor fields $\Omega_i$, which can be found by
expressing the parameter $\eta_i$ of the S-supersymmetry
transformation in terms of the parameters of the Q-supersymmetry
transformations, i.e.,
\begin{equation}
  \label{eq:s-susy-vector}
  \hat\eta_i= - X^{-1}  \big[ \Slash{\mathcal{D}} X\epsilon_i
     +\ft14 \varepsilon_{ij}  \hat F_{bc}^-
   \gamma^{bc}\epsilon^j +\tfrac12 Y_{ij} \epsilon^j\big] \,.
\end{equation}
Here we have replaced the supercovariant derivative $D_a$ by a
derivative $\mathcal{D}_a$, which is covariant with respect to only
the linearly realized bosonic symmetries. We should stress here that
special conformal boosts are not realized linearly. Usually this does
not lead to additional terms when considering derivatives on quantities
that themselves are invariant under these boosts. To avoid
confusion we will usually write the conformal gauge connection
$f_\mu{}^a$ explicitly in the purely bosonic expressions and not keep
it implicit as we do when dealing with supercovariant derivatives.

In this strategy the initial vector multiplet plays a key role, but in
due course we will demonstrate that the results will be independent of
the choice of the particular supermultiplet from where one starts this
procedure. We should also mention that all the constraints can
alternatively be obtained by exploiting the observation given below
\eqref{eq:ratio-chiral}. Namely, one can start from bosonic
expressions constructed from various supermultiplet components that
are invariant under dilatations and chiral transformations, and
explore the fact that they must vanish under repeated supersymmetry
transformations. We shall comment on this aspect when considering the
specific results of our calculations.

As explained earlier we subsequently require that all supercovariant
fermionic quantities vanish under supersymmetry and so must their
supersymmetry variations. Hence the superconformal derivative
$D_a\Omega_i$ is assumed to vanish identically. What remains is to
ensure that also its variation will vanish under the particular
combination of Q- and S-supersymmetry defined by
\eqref{eq:s-susy-vector}. To investigate the invariance of
$D_a\Omega_i$, let us first define the superconformal derivative,
\begin{equation}
  \label{eq:Da-Omega}
  D_a\Omega_i= \mathcal{D}_a\Omega_i -\Slash{D} X \psi_{a i}
     -\ft14 \varepsilon_{ij}  \hat F^-{\!\!\!}_{bc} \,
   \gamma^{bc}\psi_a{}^j -\tfrac12 Y_{ij} \psi_a{}^j - X\,\phi_{a
     i} \,, 
\end{equation}
where $\psi_\mu{}^i$ and $\psi_{\mu i}$ denote the chiral and
anti-chiral components of the gravitino field that is the gauge field
associated with Q-supersymmetry. The gauge fields of S-supersymmetry
are not elementary but composite fields denoted by $\phi_{\mu i}$ and
$\phi_{\mu}{}^{i}$. Its explicit definition can be found in
e.g. \cite{deWit:2010za,Butter:2013lta}.  The derivative
$\mathcal{D}_\mu$ is covariant under all the linearly acting bosonic
transformations, namely dilatations, local Lorentz transformations and
local R-symmetry transformations. Since we assumed that the fermionic
gauge field must also vanish in the supersymmetric limit we indeed
have $D_a\Omega_i=0$.

Now consider the supersymmetry variation of $D_a\Omega_i$, restricting
ourselves to the purely bosonic terms, using that the generic supersymmetry
variations of the Q- and S-supersymmetry gauge fields are given  (up
to terms proportional to fermionic bilinears) by 
\begin{align}
  \label{eq:psi-phi}
  \delta \psi_{\mu}{}^{i} =&\, 2 \,\mathcal{D}_\mu \epsilon^i - \tfrac{1}{8}
  T_{ab}{}^{ij} \gamma^{ab}\gamma_\mu \epsilon_j - \gamma_\mu \eta^i
  \,, \nonumber \\[2mm]
   \delta\phi_\mu{}^i =&\, - 2\,f_\mu{}^a\gamma_a\epsilon^i + \ft14
  R({\cal V})_{ab}{}^{\, i}{}_{\!j}  \gamma^{ab}\gamma_\mu \epsilon^j
  + \tfrac1{2}\mathrm{i}  R(A)_{ab}\gamma^{ab}\gamma_\mu \epsilon^i
  -\tfrac1{8} \Slash{D} T^{ab\,ij} \gamma_{ab} \gamma_\mu \epsilon_j
  + 2\,\mathcal {D}_\mu\eta^i   \,,
\end{align}
where $f_\mu{}^a$ is the gauge  field of special conformal boosts,
which is a composite field whose bosonic terms take the form
\begin{equation}
  \label{eq:conf-gauge-field}
      f_\mu{}^a= \tfrac12 R(\omega,e)_\mu{}^a -\tfrac14 \big(D+\tfrac13
  R(\omega,e)\big) e_\mu{}^a -\tfrac12\mathrm{i}\tilde R(A)_\mu{}^a +
  \tfrac1{16} T_{\mu b} {}^{ij} T^{ab}{}_{ij} \,.
\end{equation}
Here $R(\omega,e)_\mu{}^a$ and $R(\omega,e)$ are the contractions of
the curvature tensor associated with the spin connection field
$\omega_\mu{}^{ab}$, defined by $R(\omega)_{\mu\nu}{}^{ab} = 2
\,\partial_{[\mu} \omega_{\nu]}{}^{ab} - 2\, \omega_{[\mu}{}^{ac}
\omega_{\nu]c}{}^b$. Furthermore $\chi^i$ and $D$ are a spinor doublet
and a real scalar field belonging to the Weyl multiplet, while
$R(A)_{\mu\nu}$ and $R(\mathcal{V})_{\mu\nu}{}^i{}_j$ denote the
curvature tensors associated with the connections of the
$\mathrm{U}(1)$ and $\mathrm{SU}(2)$ R-symmetry, respectively.

Of course, for consistency one must also determine the constraints
from full supersymmetry on the conformal supergravity background. As a
first step in that direction we will therefore also include the
consequences of the supersymmetry invariance of the spinor $\chi^i$,
which belongs to the Weyl multiplet. An independent analysis of the
supersymmetry conditions based only on the Weyl multiplet fields will
be discussed at the end of this section. Under supersymmetry $\chi^i$
transforms as follows,
\begin{equation}
  \label{eq:delta-chi}
    \delta \chi^i = - \tfrac{1}{12} \gamma^{ab} \, \Slash{D} T_{ab}{}^{ij}
  \, \epsilon_j + \tfrac{1}{6} R(\mathcal{V})_{\mu\nu}{}^i{}_j
  \gamma^{\mu\nu} \epsilon^j -
  \tfrac{1}{3} \mathrm{i} R(A)_{\mu\nu} \gamma^{\mu\nu} \epsilon^i + D
  \epsilon^i +
  \tfrac{1}{12} \gamma_{ab} T^{ab ij} \eta_j \, .  
\end{equation}
In evaluating the consequences of the above results one may assume
that both $X$ and $T_{ab}{}^{ij}$ are non-vanishing. The reason is
that they are the lowest-weight fields of the two multiplets, so that
their vanishing would imply that the corresponding multiplets will vanish. 

Upon substituting \eqref{eq:s-susy-vector} it turns out that
$\delta(D_a\Omega_i) = 0$ and $\delta\chi^i=0$ give rise to the
following conditions,
\begin{align}
  \label{eq:full-susy-vector-Weyl}
  R(\mathcal{V})_{\mu\nu} {}^i{}_j=&\,R(A)_{\mu\nu} = R(D)_{\mu\nu} =Y_{ij}
  =0\,,  \nonumber \\[1mm]
   D=&\, \tfrac1{48} \big[ X^{-1}\, \varepsilon_{ij}T_{ab}{}^{ij}\, \hat
  F^{-ab} + \bar X^{-1} \,\varepsilon^{ij}T_{ab ij}\, \hat
  F^{+ab}\big]
  \,,  \nonumber \\[1mm]
  \hat F^-{\!\!}_a{}^c\,T_{cb}{}^{ij} =&\, T_{ac}{}^{ij} \,\hat
  F^{-c}{}_{b}\, \,, \nonumber  \\[1mm]
  \bar X\,\varepsilon_{ij}\,T_{ab}{}^{ij}\, \hat F^{-ab} =&\, X\,
  \varepsilon^{ij} \,T_{ab ij}\, \hat F^{+ ab}\,. 
\end{align} 
The third equation implies that $\hat F^{-ab}$ is proportional to
$\bar X\,\varepsilon_{ij}\,T_{ab}{}^{ij}$, with a proportionality
factor that is invariant under local dilatations and $\mathrm{U}(1)$
R-symmetry transformations. Using also the the second and fourth
equation in \eqref{eq:full-susy-vector-Weyl}, one can determine this
factor and obtain the relation
\begin{equation}
  \label{eq:F-minus-T}
  \hat F_{ab}^- = \frac{24 \,D   \, X\, T_{ab}{}^{ij} \,\varepsilon_{ij}}
  {(T^{cdkl} \,\varepsilon_{kl} )^2} \,.
\end{equation}
Here we have assumed that $T_{ab}{}^{ij}$ is not null, that is,
$(T_{ab}{}^{ij} \varepsilon_{ij})^2 \neq 0$. We will continue making
this assumption from now on.\footnote{ 
  The case where $(T_{ab}{}^{ij} \varepsilon_{ij})^2$ vanishes (in
  spite of the fact that $T_{ab}{}^{ij}\not=0$) is rather special but
  can still be dealt with by using the same method. Since the results are
  not substantially different, we ignore this case here. } 

Furthermore we also derive the following conditions involving
derivatives,
\begin{align}
    \label{eq:evaluation-full-susy-vector-weyl-2}
   \mathcal{D}_a \big(X\, T^{abij}) =&\, 0\,, \nonumber\\[1mm]
   \mathcal{D}_a \big(X\, T^{ab}{}_{ij} ) =&\,
  2\,\varepsilon_{ij}\,\mathcal{D}_a \hat F^{-ab} \,, \nonumber\\[1mm]
   \mathcal{D}_a \hat F^{-ab}  =&\,   - \mathcal{D}_a\ln(X/\bar
  X)\, \hat F^{-ab} \,, \nonumber \\[1mm]
  \mathcal{D}_a \hat F^{-bc} - \mathcal{D}_a \ln X\,\hat F^{-bc}
  =&\,  -2\,\big[ \mathcal{D}^{[b} \ln(X\bar X)\,  \hat F^{-c]}{}_a -
  \mathcal{D}_d\ln(X/\bar X) \,\hat F^{-d[b} 
  \,\delta^{c]}{\!}_a\,\big]^{[bc]-} \,,    \nonumber \\[1mm]
   X\, D_{(a} D_{b)} X - 2\, \mathcal{D}_aX\,\mathcal{D}_bX = &\,
      \frac{X}{2\,\bar X} \,\hat F^-{\!\!\!}_{a}{}^c \hat F^+{\!\!\!}_{cb} 
   - \frac12 \eta_{ab} \Big[ (\mathcal{D}_cX)^2  +\frac1{16} X\,\hat
   F^{-cd}\,T_{cd}{}^{ ij} \varepsilon_{ij}\,\Big] \,, 
\end{align}
where, in the last equation, $D_{(a}D_{b)} X
\equiv\big(\mathcal{D}_{(a} \,\mathcal{D}_{b)} +\,f_{\mu (a}\,
e_{b)}{}^\mu\, \big)X$. This equation thus leads to a condition on the
field $f_\mu{}^a$ and therefore on $R(\omega,e)_\mu{}^a$.  The
imaginary part of the second equation is consistent with the Bianchi
identity on the field strength associated with the vector gauge field
$W_\mu$. The last term in the fourth equation
\eqref{eq:full-susy-vector-Weyl} involves an anti-selfdual projection
on the indices $[bc]$. When this is taken into account, the result
takes the form
\begin{equation}
  \label{eq:evaluation-full-susy-vector-weyl-3}
  \mathcal{D}_a \hat F^{-bc} - \mathcal{D}_a \ln (X\bar X) \,\hat F^{-bc}
   +2\,\mathcal{D}^{[b} \ln X\,  \hat F^{-c]}{}_a 
   -2\, \mathcal{D}_d \ln X\,  \hat F^{-d[b} \,\delta_a{\!}^{c]}
   =0\,,
\end{equation}
which is conformally invariant in agreement with our original assumption.

We note one more equation that follows from the first three equations of
\eqref{eq:evaluation-full-susy-vector-weyl-2}, namely 
\begin{equation}
  \label{eq:calA-constraint}
  \big(\hat F^{-ab} + \tfrac14 X\,T^{ab}{}_{ij} \varepsilon^{ij} \big)
  \mathcal{A}_b =0\,,
\end{equation}
where 
\begin{align}
  \label{eq:def-calA}
  \mathcal{A}_\mu \equiv -\tfrac12 \mathrm{i} \mathcal{D}_\mu
  \ln[X/\bar X] =  A_\mu  -\tfrac12 \mathrm{i} \partial_\mu
  \ln[X/\bar X] 
\end{align}
Obviously $\mathcal{A}_\mu$ is invariant under chiral $\mathrm{U}(1)$
and dilatations. Because $R(A)_{\mu\nu}=0$ it follows that
$\partial_{[\mu}\mathcal{A}_{\nu]}=0$. Substituting
\eqref{eq:F-minus-T} into \eqref{eq:calA-constraint}, one derives,
after multiplication with the selfdual tensor $T_{abij}$ and making
use of the standard identities for products of (anti-)selfdual
tensors,
\begin{equation}
  \label{eq:TT-96-1}
  \big[\varepsilon^{ij} T_{ab \,ij} \,T^{ac kl}\varepsilon_{kl}  + 24\, D\, \delta_b{}^c \big]\,
  \mathcal{A}_c  = 0\,, 
\end{equation}
The first term in this equation contains the product of a selfdual and
an anti-selfdual tensor which is symmetric and traceless, and whose
square must be proportional to the identity matrix. In this way one
can obtain the following equation,
\begin{equation}
  \label{eq:special-D-value}
  \left( \frac{D^2}{\big\vert(T^{abij}\varepsilon_{ij})^2\big\vert^2} - \frac1{(96)^2}
    \right)\,\mathcal{A}_\mu  = 0\,. 
\end{equation}

At this point we have not yet evaluated all the constraints of full
supersymmetry on the Weyl multiplet. Besides the spinor field $\chi^i$
that we have already considered, there exists a supercovariant
tensor-spinor, $R(Q)_{ab}{}^i$, which is the superconformal field
strength of the gravitini fields. It emerges as the supersymmetry
variation of the tensor field $T^{abij}$, so that it must
vanish. Under Q- and S-supersymmetry $R(Q)_{ab}{}^i$ transforms as
\begin{equation}
  \label{eq:delta-R(Q)}
  \delta R(Q)_{ab}{}^i = -\tfrac12 \Slash{D}T_{ab}{}^{ij} \epsilon_j +
  R(\mathcal{V})^{-}_{ab}{}^i{}_j \, \epsilon^j -\tfrac12
    \mathcal{R}(M)_{ab}{}^{cd} \,\gamma_{cd} \epsilon^i +\tfrac18
  T_{cd}{}^{ij} \gamma^{cd}\gamma_{ab} \,\eta_j \,,
\end{equation}
where $\mathcal{R}(M)_{ab}{}^{cd}$ is a modification of the curvature
associated with the spin connection field $\omega_\mu{}^{ab}$. 

Requiring $\delta R(Q)_{ab}{}^i=0$, and using again
\eqref{eq:s-susy-vector}, leads to two more equations,
\begin{align}
      \mathcal{D}_a T^{bcij}   -\mathcal{D}_a\ln X\,  T^{bcij} +
  2\,\mathcal{D}^{[b} \ln X\,T^{c]}{}_a{}^{ij} - 2\,
  \mathcal{D}_d\ln X\, T^{d [b ij}\, \delta^{c]}{\!}_a
  =0\,, \nonumber\\
  \mathcal{R}(M)^-_{ab\,cd} - \frac1{2\,\vert X\vert^2} \,
  (\varepsilon_{ij} \bar X\,T_{a[c}{}^{ij} ) \, \hat F^-{\!\!}_{d]b} \big\vert^{[ab]-}
  = 0\,.
  \label{eq:RMDT}
\end{align}
From the first equation we derive 
\begin{equation}
  \label{eq:DT-T2}
  \varepsilon_{kl} T_{ab}{}^{kl} \,\mathcal{D}_c T^{cbij} \varepsilon_{ij} = -\tfrac18
  \mathcal{D}_a (T^{bckl}\varepsilon_{kl})^2 \,,
\end{equation}
by making use of the identities that hold for contractions of
(anti-)selfdual tensors. Furthermore one derives, upon combining
\eqref{eq:F-minus-T}, \eqref{eq:evaluation-full-susy-vector-weyl-3}
and the first equation of \eqref{eq:RMDT}, that certain ratios of
fields must be constant,
\begin{equation}
  \label{eq:constant-propto}
  \frac{X^2}{(T^{abij}\varepsilon_{ij})^2} = \mbox{constant}
    \,,\qquad \frac{D}{\big\vert(T^{abij}\varepsilon_{ij})^2\big\vert} = \mbox{constant}\,.  
\end{equation}
These expressions can be regarded as the lowest-weight components
of a chiral or real supermultiplet, respectively, with
$w=c=0$. According to the theorem discussed earlier in this section,
such multiplets must indeed be equal to a constant in the
supersymmetric limit. This observation enables an alternative
derivation of the same results that we are deriving in this section.

The second equation~\eqref{eq:RMDT} involves an anti-selfdual
projection over the the index pair $[ab]$ (because of the symmetry of
this term, it is also anti-selfdual in $[cd]$), while
$\mathcal{R}(M)^-_{ab\,cd}$ is anti-selfdual in both index pairs
$[ab]$ and $[cd]$. Using \eqref{eq:F-minus-T} the equation then takes
the form
\begin{equation}
  \label{eq:M-T-D}
   \mathcal{R}(M)^-_{ab\,cd} - \frac{12\,D} {(T^{abij}
     \varepsilon_{ij})^2}\, P^-_{ab,cd} = 0\,, 
\end{equation}
where\footnote{
  Note that we are using Pauli-K\"all\'en conventions so that the Levi-Civita
  symbol is effectively pseudo-real.} 
\begin{equation}
  \label{eq:def-proj}
  P_{ab,cd}^- \equiv  T_{a[c} \,T_{d]b}\big\vert^{[ab]-}=  \tfrac18
  \big(\delta_{a[c}\,\delta_{d]b} -\tfrac12\varepsilon_{abcd}\big) {(T^{efij}
     \varepsilon_{ij})^2}  - \tfrac12 \varepsilon_{ij} \, 
  T_{cd}{}^{ij} \,T_{ab}{}^{kl} \,\varepsilon_{kl}   \,. 
\end{equation}

By now we have obtained a number of conditions that do not explicitly
involve the vector multiplet fields. A relevant question is therefore
whether the Weyl multiplet alone (i.e. without being coupled to a
vector multiplet) requires the same conditions when imposing
supersymmetry. Therefore we repeat the same procedure but now without
coupling to a vector multiplet. Hence we start with the supersymmetry
variation of the field $\chi^i$ shown in \eqref{eq:delta-chi}, and
choose $\hat\eta_i$ such that its supersymmetry variation
vanishes.

At this point the reader may wonder whether a different choice for
$\hat\eta_i$ would not affect the results of the previous analysis, so
that they would become incompatible with the new ones that we are
about to derive. This is actually not the case, as one can simply see
by considering the supersymmetry variation of the S-supersymmetric
linear combination, $T^{abij} \gamma_{ab} \Omega_j -24\, X\,\chi^i$,
whose vanishing under Q-supersymmetry is obviously independent of
whether $\hat\eta_i$ is chosen such that $\delta\Omega_i$ or
$\delta\chi^i$ will vanish. To base the analysis on S-supersymmetric
combinations of spinors was precisely the approach followed in
\cite{LopesCardoso:2000qm}. Hence it follows that the choice of
$\hat\eta_i$ is irrelevant, and it is again obvious that the fermionic
gauge orbit associated with S-supersymmetry is not affected, as was
emphasized earlier. Our approach of adopting a specific $\hat\eta_i$
associated with a specific supermultiplet is thus a matter of
convenience when considering separate configurations of
supermultiplets.

Using the expression for $\hat\eta_i$ that is found by solving
$\delta \chi^i = 0$ directly, one can evaluate
the variations of $D_a\chi^i$ and $R(Q)^i_{ab}$, requiring them to
vanish also. This calculation is completely similar to the approach
followed before. A careful evaluation then shows that all the
constraints of the Weyl multiplet imposed by requiring supersymmetry
coincide fully with the constraints that we have evaluated before,
starting from the vector multiplet (possibly exploiting the first
equation of \eqref{eq:constant-propto}). 

Let us now return the last equation of
\eqref{eq:evaluation-full-susy-vector-weyl-2}, which involves terms
quadratic in derivatives and yields an expression for the composite
connection $f_\mu{}^a$ associated with the conformal boosts,
\begin{align}
  \label{eq:f-value-2}
  f_a{}^b =&\, -\mathcal{D}_a\mathcal{D}^b\ln X + \mathcal{D}_a\ln X\,
  \mathcal{D}^b\ln X -\tfrac12\delta_a{}^b \,\big(\mathcal{D}_c\ln
  X\big)^2 -\tfrac34 \delta_a{}^b D \nonumber\\
  &\, - \frac{288\,D^2
    \,\varepsilon_{ij} T_{ac}{}^{ij}\,T^{bc}{}_{kl}\varepsilon^{kl} }
  {\big\vert(T^{demn}\varepsilon_{mn})^2\big\vert^2 }
\end{align}
Whereas the left-hand side is manifestly real, the right-hand side is
not. To analyze this we note that $\mathcal{D}_\mu X = \mathcal{D}_\mu
\vert X\vert +\mathrm{i} \mathcal{A}_\mu$, where $\mathcal{A}_\mu$ has
been defined in \eqref{eq:def-calA}.  The reality of
\eqref{eq:f-value-2} then implies 
\begin{equation}
  \label{eq:D-cA-constr}
  \mathcal{D}_a\mathcal{A}_b - 2\,\mathcal{A}_{(a} \,\mathcal{D}_{b)}\ln\vert X\vert
  -\eta_{ab} \,\mathcal{A}_c\,\mathcal{D}^c\ln\vert X\vert = 0\,, 
\end{equation}
where we note that \eqref{eq:special-D-value} implies that
$\mathcal{A}_\mu=0$ for $\vert D\vert \not=
\tfrac1{96}\,\vert(T^{abij}\varepsilon_{ij})^2\vert$.  Hence we obtain
the following form for the real part of \eqref{eq:f-value-2}
\begin{align}
  \label{eq:f-value-3}
  f_a{}^b =&\, -\mathcal{D}_a\mathcal{D}^b\ln \vert X\vert +
  \mathcal{D}_a\ln\vert X\vert\,
  \mathcal{D}^b\ln\vert X\vert - \mathcal{A}_a \,
  \mathcal{A}^b \nonumber\\
  &\, -\tfrac12\delta_a{}^b \,\big[ \big(\mathcal{D}_c\ln 
  \vert X\vert\big)^2 -\mathcal{A}^c \mathcal{A}_c + \tfrac32  D\big]
   - \frac{288\,D^2
    \,\varepsilon_{ij} T_{ac}{}^{ij}\,T^{bc}{}_{kl}\varepsilon^{kl} }
  {\big\vert(T^{demn}\varepsilon_{mn})^2\big\vert^2 }
\end{align}

This completes the derivation of a consistent set of covariant
equations that characterize the fully supersymmetric configurations
consisting of a vector and the Weyl supermultiplet. What remains is to
present the results for the components of the Riemann tensor. Up to
this point we have fully preserved the covariance with respect to the
bosonic symmetries of the superconformal group, so that the
spin-connection field $\omega_\mu{}^{ab}$ depends both on the
vierbein $e_\mu{}^a$ and on the dilatational gauge field $b_\mu$. Hence the
associated curvature $R(\omega)_{\mu\nu}{}^{ab}$ is only identical to
the Riemann tensor when $b_\mu$ vanishes. For a conformally invariant
action $b_\mu$ will be absent, while otherwise one still has the
option to impose $b_\mu=0$ as a gauge condition.  Comparing
\eqref{eq:f-value-3} to \eqref{eq:conf-gauge-field}, one derives the
following expression for the Ricci tensor and scalar,
\begin{align}
  \label{eq:ricci}
  \mathcal{R}(\omega,e)_{ab} =&\, -2\,\mathcal{D}_a\mathcal{D}_b\ln\vert X\vert +
  2\,\mathcal{D}_a\ln\vert X\vert\, \mathcal{D}_b\ln\vert X\vert
  -2\, \mathcal{A}_a\,\mathcal{A}_b \nonumber \\
  &\, -\eta_{ab} \Big[
  \mathcal{D}^c\mathcal{D}_c\ln\vert X\vert + 2\,\big(\mathcal{D}_c\ln
  \vert X\vert\big)^2  + 2\,\mathcal{A}^c \mathcal{A}_c +3\, D\Big]  \nonumber\\
  &\, - \Big[\frac{1}{16} + \frac{576\,D^2 }
   {\big\vert(T^{demn}\varepsilon_{mn})^2\big\vert^2} \Big]
   \,\varepsilon_{ij} T_{ac}{}^{ij}\,T_{b}{}^{c kl}\varepsilon_{kl}
   \,. \nonumber\\
   \mathcal{R}(\omega,e) =&\, -6\,\mathcal{D}^a\mathcal{D}_a\ln\vert X\vert -
  6\,\mathcal{D}^a\ln\vert X\vert\, \mathcal{D}_a\ln\vert X\vert
  +6\, \mathcal{A}^2 -12\, D  \,.
\end{align}
Note that the Ricci tensor is in general not symmetric in the presence
of the field $b_\mu$. Finally we note that
\begin{equation}
  \label{eq:cR(M)}
  \mathcal{R}(M)_{ab}{}^{cd}  =\mathcal{C}(e, \omega) _{ab}{}^{cd}
  +D\,\delta_{ab}{}^{cd} + \cdots\,,
\end{equation}
where the suppressed terms are proportional to
$R(A)_{\mu\nu}$ and to fermion bilinears, which all vanish in the
supersymmetric background. Making use of \eqref{eq:M-T-D} one then
derives the expression for the Weyl tensor, 
\begin{equation}
  \label{eq:R(M)}
  \mathcal{C}(e,\omega)_{ab}{}^{cd}  = D \,\left[2\,
    \delta_{ab}{}^{cd}   
	- \frac{6 \,\varepsilon_{ij} T_{ab}^{ij}\, T^{cd kl} \varepsilon_{kl}}
        {(\varepsilon_{mn} T^{demn})^2}
	- \frac{6 \,\varepsilon^{ij} T_{abij}\, T^{cd}{}_{kl} \varepsilon^{kl}}
        {(\varepsilon^{mn} T^{de}{}_{mn})^2}	\right]~.
\end{equation}

\section{Three other short multiplets}
\label{sec:short-mutiplets}
\setcounter{equation}{0}
In this section, we consider the remaining $N=2$ short multiplets
commonly encountered. They are the tensor multiplet, the non-linear
multiplet, and the (on-shell) hypermultiplet. Their distinctive
feature is that their lowest-weight components are scalar fields
transforming under the $\mathrm{SU}(2)$ R-symmetry. For the tensor
multiplet these fields are the pseudo-real $\mathrm{SU}(2)$ vector
$L_{ij}$, for the non-linear multiplet it is given by a space-time
dependent $\mathrm{SU}(2)$ element $\Phi^i{}_\alpha$, and for the
hypermultiplet they are represented by certain sections
$A(\phi)_i{}^\alpha$ of a hyperk\"ahler cone.\footnote{The indices
$\alpha$ for the non-linear multiplet and the hypermultiplet sections
are unrelated. For example, the former take the values $\alpha=1,2$
while the latter take the values $\alpha=1,\cdots,r$.}
These quantities will be
introduced shortly. We assume that their $\mathrm{SU}(2)$ invariant
norms are non-vanishing. For the non-linear multiplet, the norm equals
$\det[\Phi^i{}_\alpha]=1$; for the tensor and the hypermultiplet, these
norms are the length $L$ of the vector $L_{ij}$ and the so-called
hyperk\"ahler potential $\chi(\phi)$, respectively, which both have
$w=2$. Their precise definitions will be given shortly.

Requiring that the scalars are invariant under supersymmetry leads to
the condition that the fermion fields must vanish.  We discover that
the presence of $\mathrm{SU}(2)$ indices on the lowest-dimension
scalars generically leads to stronger conditions on the Weyl multiplet
than the ones found for the vector multiplet in the previous
section. Since all the underlying principles of the analysis have
already been exhibited in the previous section, we keep the
presentation rather concise.  Obviously the conditions on the Weyl
multiplet alone may be assumed.  In particular, taking
$R(\mathcal{V})_{\mu\nu} {}^i{}_j=R(A)_{\mu\nu} = R(D)_{\mu\nu} =0$
from the start will simplify the analysis.  An important condition,
which will play a key role in many of the formulae, is
\begin{align}\label{eq:DerScale}
\mathcal{D}_a \ln \big\vert(T_{bc}{}^{ij} \veps_{ij})^2 \big\vert 
= \cD_a \ln (X \bar X) =
\begin{cases}
\mathcal{D}_a \ln L~, & \text{tensor multiplet} \\
-V_a~, & \text{non-linear multiplet} \\
\mathcal{D}_a \ln \chi~,& \text{hypermultiplet}
\end{cases}
\end{align}
where $V_a$ is a vector component of the non-linear multiplet, and $L$
and $\chi$ are the two composite real $w=2$ scalar fields introduced
above.  These conditions are consistent with the (now familiar)
observation that any $w=c=0$ scalar field must be constant, and so
$|(T_{ab}{}^{ij} \veps_{ij})^2|$ must be proportional to $X\bar X$,
$L$ and $\chi$ for a vector multiplet, tensor multiplet and
hypermultiplet, respectively. Note that the vector $V_a$ is not
invariant under special conformal boosts.

In contrast with the previous section, we will find that for the three
multiplets discussed here, the $w=2$ scalar field $D$ of the Weyl
multiplet will be required to vanish.  This turns out to have major
consequences for both the Weyl multiplet and for any vector multiplet.
Invoking \eqref{eq:F-minus-T} and \eqref{eq:M-T-D}, one derives the
following constraints on the Weyl multiplet and any vector
multiplet:
\begin{equation}
  \label{eq:add-constr-Weyl}
  D=0 \quad \implies \quad \mathcal{R}(M)_{ab\,cd}=0\,, \qquad \hat F_{ab}= 0\,. 
\end{equation}
The second equation implies that the Weyl tensor must vanish as a
result of \eqref{eq:R(M)}.
The third equation of \eqref{eq:add-constr-Weyl} leads to a constraint on the 
vector multiplet field strength,
\begin{equation}
  \label{eq:vector-field-strength-cond}
  F_{\mu\nu} \equiv 2\, \partial_{[\mu} W_{\nu]} =
  \tfrac14\big[ X \,T_{\mu\nu\,ij}\,\varepsilon^{ij}  +\bar X
  \,T_{\mu\nu}{}^{ij}\,\varepsilon_{ij} \big] \,.
\end{equation}
Another consequence of $D=0$ is given by \eqref{eq:special-D-value}, which implies that
\begin{align} \label{eq:add-constr-phase}
  \mathcal{A}_\mu = -\tfrac{1}{2} \ri \mathcal{D}_\mu \ln (X / \bar X)
	= -\tfrac{1}{4} \mathrm{i} \mathcal{D}_\mu \ln \big[(T_{bc}{}^{ij}
        \veps_{ij})^2 / (T^{de}{}_{kl} \veps^{kl})^2\big] = 0~.
\end{align}
This determines the U(1) gauge connection in terms of the phase of
$T_{ab}{}^{ij}$ (or $X$).
The final two conditions we will encounter are the analogues of
\eqref{eq:RMDT} and \eqref{eq:f-value-3}, found by making the replacement
\eqref{eq:DerScale} with the additional constraints
\eqref{eq:add-constr-Weyl} and \eqref{eq:add-constr-phase}.

\subsection{The tensor multiplet}
The tensor multiplet consists of a pseudo-real $\mathrm{SU}(2)$
triplet of scalar fields $L_{ij}$, which has Weyl weight $w=2$ and
satisfies the pseudo-reality constraint $(L^{ij})^* =
\varepsilon_{ik}\varepsilon_{jl} L^{kl}$, a doublet of spinors
$\varphi^i$, a two-form gauge field $E_{\mu \nu}$, and a complex
scalar $G$. Their Q- and S-supersymmetry transformations are
\begin{align}
  \label{eq:tensor-tr}
  \delta L_{ij} =& \,2\,\bar\epsilon_{(i}\varphi_{j)} +2
  \,\varepsilon_{ik}\varepsilon_{jl}\,
  \bar\epsilon^{(k}\varphi^{l)}  \,,\nonumber \\  
  \delta\varphi^{i} =& \,\Slash{D} L^{ij} \,\epsilon_j +
  \varepsilon^{ij}\,\Slash{\hat E}^I \,\epsilon_j - G \,\epsilon^i
  + 2 L^{ij}\, \eta_j \,,\nonumber \\  
  \delta G =& \,-2 \,  \bar\epsilon_i \Slash{D} \, \varphi^{i} \,
  - \bar\epsilon_i  ( 6 \, L^{ij} \, \chi_j + \tfrac1{4} \,
    \gamma^{ab}  T_{ab jk} \, \varphi^l \,
    \varepsilon^{ij} \varepsilon^{kl}) + 2 \, \bar{\eta}_i\varphi^{i}
    \, ,\nonumber \\  
  \delta E_{\mu\nu} =& \, \mathrm{i}\bar\epsilon^i\gamma_{\mu\nu} 
  \varphi^{j} \,\varepsilon_{ij} - \mathrm{i}\bar\epsilon_i\gamma_{\mu\nu}
  \varphi_{j} \,\varepsilon^{ij} \,  + \,2 \mathrm{i} \, L_{ij} \,
  \varepsilon^{jk} \, \bar{\epsilon}^i \gamma_{[\mu} \psi_{\nu ]k}  
  - 2 \mathrm{i}\,  L^{ij} \, \varepsilon_{jk} \, \bar{\epsilon}_i
    \gamma_{[\mu} \psi_{\nu ]}{}^k \, ,
\end{align}
where $D_a$ are the superconformally covariant derivatives, and $\hat
E^a$ equals the dual of a supercovariant three-form field strength,
\begin{align}
  \hat E^\mu =&\, \tfrac{1}{2}\mathrm{i}\, e^{-1} \, \varepsilon^{\mu \nu
  \rho \sigma}  \Big[\partial_\nu E_{\rho \sigma}   
  - \tfrac{1}{2} \mathrm{i} \bar{\psi}{}^i_\nu  \gamma_{\rho\sigma}
  \varphi^j \varepsilon_{ij} 
  + \tfrac{1}{2}\mathrm{i} \bar{\psi}_{\nu i} \gamma_{\rho\sigma}
  \varphi_{j} \varepsilon^{ij}  
  -\mathrm{i} \, L_{ij}   \varepsilon^{jk} 
  \bar{\psi}_\nu{}^i \gamma_\rho \psi_{\sigma k}\Big] \,.
\end{align}
A supersymmetric field configuration for this multiplet can be found by
following the same steps as for the vector multiplet. We note the
convenient identity, $L^{ij}L_{jk} = \delta^i{}_k\,L^2$, where the
modulus $L$ of the $\mathrm{SU}(2)$ triplet is given by $L^2 =
\tfrac{1}{2} L^{ij} L_{ij}$. We will assume that $L$ is non-vanishing
and impose $\delta \varphi^i=0$ by choosing
\begin{align}
  \label{eq:eta-tensor} 
  \hat\eta_i = - \tfrac12 L_{ij} \,L^{-2} \big[ \Slash{\mathcal{D}} L^{jk}
  \,\epsilon_k + \varepsilon^{jk} {\Slash{\hat E}} \,\eps_k - G\, 
  \epsilon^j \big]~,
\end{align}
where all terms containing fermionic bilinears can be dropped.  Next,
we impose the conditions $\delta (D_a\varphi^i)=0$ and $\delta \chi^i=
\delta R(Q)_{ab}{}^i = 0$ and analyze their consequences. Although the
latter two conditions have already been investigated separately, it
turns out that when combining these with the condition $\delta
(D_a\varphi^i) =0$, while using the expression \eqref{eq:eta-tensor}, one
more readily obtains the results \eqref{eq:add-constr-Weyl}, strongly
restricting the Weyl multiplet. Assuming as before that
$T_{ab}{}^{ij}$ does not vanish leads to the conditions
\begin{align}
  \label{eq:TensorGij}
  G = \hat E_a = 0~, \qquad
  L_{ik} \overset{\leftrightarrow}{\mathcal{D}}_a L^{kj} = 0~,
\end{align}
which force the two-form $E_{\mu\nu}$ to be pure gauge and restrict
$\mathcal{D}_a L_{ij} = L_{ij} \,\mathcal{D}_a \ln L$, or
\begin{equation}
  \label{eq:L-L-1-constant}
  \mathcal{D}_a (L_{ij}\,L^{-1}) = 0\,. 
\end{equation}
We find that the derivative of $T_{ab}{}^{ij}$ is given by
\eqref{eq:RMDT} with the replacement $\mathcal{D}_a \ln X
\rightarrow \tfrac{1}{2} \mathcal{D}_a \ln L$, implying both
\eqref{eq:add-constr-phase} and \eqref{eq:DerScale}.  Similarly, the
analogue of \eqref{eq:f-value-3} is reproduced.

\subsection{The non-linear multiplet}
Next we consider the case of the `non-linear multiplet' in a conformal
supergravity background \cite{deWit:1980ib,deWit:1980tn}.  This
multiplet consists of a scalar $\mathrm{SU}(2)$ matrix
$\Phi^i{}_\alpha$ with $\alpha=1,2$, a fermion doublet with negative
(positive) chirality components $\lambda^i$ ($\lambda_i$), a complex
anti-symmetric tensor $M^{ij}$ and a real vector field $V^a$.  Because
$\Phi^i{}_\alpha$ is an element of SU(2), it must have
vanishing Weyl weight and its inverse matrix is given by its hermitian
conjugate denoted by $\Phi^\alpha{}_i$.  Under Q- and S-supersymmetry,
the fields transform as
\begin{align}
  \label{eq:nonlinear-mult-transf}
  \delta \Phi^i{}_\alpha &= (2 \,\bar \epsilon^i \lambda_j -
  \delta^i{}_j \,\bar\epsilon^k \lambda_k - \mathrm{h.c.})
  \,\Phi^j{}_\alpha~, \nonumber\\[1mm]
  \delta\lambda^i &= -\tfrac{1}{2} \slash{V} \epsilon^i -
  \tfrac{1}{2} M^{ij} \epsilon_j + \Phi^i{}_\alpha \Slash{D} \Phi^\alpha{}_j
  \epsilon^j + \eta^i~, \nonumber\\[1mm]
  \delta M^{ij} &= 12 \,\bar\epsilon^{[i}
  \chi^{j]} + \tfrac{1}{2} \bar\epsilon^k \gamma^{ab} \lambda_k\,
  T_{ab}{}^{ij} - 4 \bar \epsilon^{[i} \slash{V} \l^{j]} - 2 \,\bar
  \epsilon^k \l_k M^{ij} + 8 \,\bar \epsilon^{[i} \big(\Slash{D} \lambda^{j]}
  + \Phi^{j]}{}_\alpha \Slash{D} \Phi^\alpha{}_k \l^k\big)~,
  \nonumber\\[1mm]
  \delta V^a &= \tfrac{3}{2} \bar \epsilon^i \gamma^a \chi_i -
  \tfrac{1}{8} \bar \epsilon^i \gamma^a \gamma^{bc} \lambda^j\, T_{bc\, ij} -
  \bar \epsilon^i \gamma^a \slash{V} \lambda_i + \bar \epsilon^i \gamma^a
  \lambda^j M_{ij} + 2\, \bar\epsilon^i \gamma^{ab} \mathcal{D} _b\lambda_i \nonumber\\
  & \qquad + 2
  \bar\epsilon_i \gamma^a \Phi^i{}_\alpha \Slash{D} \Phi^\alpha{}_j \lambda^j
  - \bar\lambda_i \gamma^a \eta^i + \mathrm{h.c.} \;, 
\end{align}
where we have suppressed terms explicitly quadratic in the fermion
fields.  In order for the supersymmetry algebra to close, the vector
$V^a$ must obey the non-linear constraint (up to terms quadratic in
the fermion fields)
\begin{align}
  \label{eq:NLConst}
  &\,D_a V^a  - \tfrac{1}{2} V^2 - 3 D - \tfrac{1}{4} M^{ij} M_{ij} +
  \mathcal{D}_a \Phi^i{}_\alpha \mathcal{D}^a \Phi^\alpha{}_i =0\,,
\end{align}
which can be interpreted as a condition on the field $D$ of the Weyl
multiplet. An unusual feature is that $V^a$ transforms under conformal
boosts, $\delta_\mathrm{K} V^a =2\,\Lambda_\mathrm{K}{}^a$. Therefore
the bosonic terms in the covariant derivative of $D_\mu V^a$ take the
form
\begin{equation}
  \label{eq:1}
  D_\mu V^a = (\partial_\mu - b_\mu) V^a  - \omega_\mu{}^{ab} \,V_b - 2\,
  f_\mu{}^a\,. 
\end{equation}
Since $V^a$ has Weyl weight $w=1$, it follows that $\delta_\mathrm{K}
(D_aV^a) = 2\,\Lambda_\mathrm{K}{}^a \,V_a$, so that the combination
$D_aV^a- \tfrac12 V^2$ is conformally invariant.

As before, the condition $\delta\lambda^i=0$ can be implemented by
making a special choice for the S-supersymmetry parameter, 
\begin{align}
  \label{eq:eta-nonlinear-mult}
  \hat\eta^i = \tfrac{1}{2} \slash{V} \epsilon^i + \tfrac{1}{2} M^{ij}
  \epsilon_j - \Phi^i{}_\alpha \Slash{\mathcal{D}} \Phi^\alpha{}_j\, \epsilon^j~.
\end{align}
Requiring $\delta (D_a\lambda^i) =0$ and $\delta\chi^i = \delta
R(Q)_{ab}{}^i = 0$ leads to a number of conditions. The Weyl multiplet
constraints are obviously implied, and one again finds
that \eqref{eq:add-constr-Weyl} should hold, along with
\begin{gather}\label{eq:NLConstraints}
M^{ij} = 0~, \qquad \Phi^i{}_\alpha \,\mathcal{D}_a \Phi^\alpha{}_j = 0~.
\end{gather}
The latter equation determines the $\mathrm{SU}(2)$ connection in terms of
$\Phi^i{}_\alpha \pa_\mu \Phi^\alpha{}_j$. In addition, one
finds
\begin{equation}
  \label{eq:cov-der-V}
   V_a = - \mathcal{D}_a
   \ln(T^{bcij}\varepsilon_{ij})^2   = - \mathcal{D}_a
   \ln(T^{bc}{}_{kl}\varepsilon^{kl})^2\,, 
\end{equation}
implying \eqref{eq:add-constr-phase} and \eqref{eq:DerScale}.
The equations \eqref{eq:M-T-D} and \eqref{eq:f-value-3},
upon replacing $\mathcal{D}_a \ln X \rightarrow - \frac{1}{2} V_a$,
are also found.

\subsection{The hypermultiplet sector}
Unlike the previous supermultiplets, hypermultiplets are realized as
an on-shell supermultiplet. Since the multiplet consists only of
scalar fields and fermions, without any gauge fields, there does not
exist a preferred basis for the fields, which are subject to non-linear redefinitions
that take the form of target-space diffeomorphisms and frame
transformations of the fermions.  For this reason, the hypermultiplets
tend to mix under supersymmetry and so it is necessary to consider the
entire hypermultiplet sector at once.

For a system of $r$ hypermultiplets, one is dealing with a $4r$-dimensional
hyperk\"ahler target space
with local coordinates $\phi^A$ and a
target-space metric $g_{AB}$, $2r$ positive-chirality spinors
$\zeta^{\bar \alpha}$ and $2r$ negative-chirality spinors
$\zeta^\alpha$. The chiral and anti-chiral spinors are related by
complex conjugation as they are Majorana spinors. They are subject to
field-dependent reparametrizations of the form $\zeta^\alpha\to
S^\alpha{}_\beta(\phi)\, \zeta^\beta$; the fields $\zeta^{\bar\alpha}$
are then redefined with the complex conjugate of
$S^\alpha{}_\beta$. The target space is subject to arbitrary
diffeomorphisms and has the standard Christoffel connection
$\Gamma_{AB}{}^C$. Likewise there exist connections
$\Gamma_A{}^\alpha{}_\beta$ and
$\Gamma_A{}^{\bar\alpha}{}_{\bar\beta}$ associated with the
field-dependent redefinitions noted above. Furthermore supersymmetry
implies the existence of an hermitian and a skew-symmetric covariantly
constant tensor, $G^{\alpha\bar\beta}$ and $\Omega^{\alpha\beta}$,
respectively. The hermitian one appears in the kinetic term for the
fermions, and the skew-symmetric one is related to
the canonical invariant antisymmetric tensor of $\mathrm{Sp}(r)$.

In order to couple the $r$ hypermultiplets to conformal supergravity,
their target-space geometry must be a $4r$-dimensional hyperk\"ahler
cone \cite{deWit:1999fp}.\footnote{ 
  Upon fixing the dilatational and $\mathrm{SU}(2)$ gauges, conformal
  supergravity is converted to Poincar\'e supergravity, and
  correspondingly the hyperk\"ahler cone is converted into a
  quaternion-K\"ahler target space \cite{deWit:1999fp,deWit:2001dj},
  in accordance with \cite{BaggerWitten}. } 
The hypermultiplet scalars transform under dilatations associated
with a homothetic Killing vector, and under the $\mathrm{SU}(2)$
R-symmetry, associated with the $\mathrm{SU}(2)$ Killing vectors of
the hyperk\"ahler cone.  The fermions transform under dilatations and
the $\mathrm{U}(1)$ factor of the R-symmetry by scale transformations
and chiral rotations, respectively.

A systematic treatment of hypermultiplets makes use of local sections
$A_i{}^\alpha(\phi)$ of an $\mathrm{Sp}(r) \times \mathrm{Sp}(1)$
bundle, where $\mathrm{Sp}(1)\cong \mathrm{SU}(2)$ refers to the
corresponding R-symmetry group. These sections transform covariantly
under R-symmetry and scale under dilatations with $w=1$. We refer to
\cite{deWit:1999fp} for further details. The Q- and S-supersymmetry
transformations on the sections and the fermions take the following
form, 
\begin{align}
  \label{eq:delta-A-hyper}
  \delta A_i{}^\alpha =&\, 2 \bar\epsilon_i \zeta^\alpha + 2 \varepsilon_{ij}
  \,G^{\alpha \bar \beta} \Omega_{\bar\beta \bar\gamma} \bar \epsilon^j
  \zeta^{\bar\gamma} - \delta_\mathrm{Q} \phi^B\,\Gamma_B{}^\alpha{}_\beta
  A_i{}^\beta~, \nonumber\\
  \delta \zeta^\alpha =&\, \Slash{D} A_i{}^\alpha
  \epsilon^i + A_i{}^\alpha \eta^i - \delta_\mathrm{Q} \phi^B\,
  \Gamma_B{}^\alpha{}_\beta \z^\beta~,
\end{align}
where $\delta_\mathrm{Q}\phi^A$ denotes the transformation rule for
the target-space scalars whose form is not relevant for what
follows. The covariant tensors $G_{\bar\alpha \beta}$ and 
$\Omega_{\bar\alpha \bar\beta}$ can be expressed as bilinears in the
covariant derivatives of the sections,
\begin{align}
  \label{eq:tensors-section}
  g^{AB}\, D_A A_i{}^\alpha \,D_B A^{j\bar\beta} =\, \delta_i{}^j\,
  G^{\alpha \bar\beta}\,, \qquad
  g^{AB}\, D_A A_i{}^\alpha \,D_B A_j{}^{\beta} =\,
  \varepsilon_{ij}\,   \Omega^{\alpha\beta} \,. 
\end{align}

A supersymmetric configuration requires that both the fermions and
their supersymmetry variations vanish.  For $r>1$, one cannot find a
choice for $\hat\eta^i$ which immediately solves $\delta \zeta^\alpha
= 0$ for all $\alpha$, so it will help to first single out one
specific fermion to solve for $\hat\eta^i$.  We will follow a similar
procedure as in \cite{LopesCardoso:2000qm} and first single out the
$w=2$ hyperk\"ahler potential $\chi$, defined by
\begin{equation}
  \label{eq:hk-potential}
  \chi = \tfrac{1}{2} \varepsilon^{ij}\,\bar \Omega_{\alpha\beta}\,
  A_i{}^\alpha\,A_j{}^\beta\,,
\end{equation}
and focus on the composite fermion $\zeta_i$ into which it varies,
\begin{align}
\delta \chi = 2 \varepsilon^{ij} \bar \epsilon_j \zeta_i + \HC~, \qquad
\zeta_i = \bar \Omega_{\alpha\beta}\,A_i{}^\alpha\,\zeta^\beta~.
\end{align}
Solving $\delta \zeta_i = 0$ leads to
\begin{align}
  \label{eq:delta-zeta-hyper}
  \hat\eta^i = \varepsilon^{ij} \, \chi^{-1} A_j{}^\beta \,\bar
  \Omega_{\beta \alpha} \Slash{D} A_k{}^\alpha \, \epsilon^k  \,.
\end{align}
Subsequently one imposes the conditions
$\delta \chi^i = \delta R(Q)_{ab}{}^i = 0$ and
$\delta (D_a \zeta_i) = 0$.
One confirms again the standard conditions on
the Weyl multiplet, including the additional conditions
\eqref{eq:add-constr-Weyl} and \eqref{eq:add-constr-phase}.
The first equation of \eqref{eq:M-T-D} and \eqref{eq:f-value-3} follow
with $\mathcal{D}_a \ln X \rightarrow \tfrac{1}{2} \mathcal{D}_a \ln \chi$.
In addition to these constraints, one finds
\begin{align}
  \label{eq:HyperGij}
  A_{(i}{}^\alpha \bar \Omega_{\alpha \beta} \mathcal{D}_a
  A_{j)}{}^\beta = 0~. 
\end{align} 
For $r>1$, one must still satisfy
$\delta \zeta^\alpha=0$. Using \eqref{eq:HyperGij}, one finds the additional
condition (trivially satisfied for $r=1$)
\begin{align}
  \label{eq:hypersection-inv}
  \mathcal{D}_a A_i{}^\alpha - \tfrac{1}{2} \mathcal{D}_a \ln \chi\, A_i{}^\alpha =
  \chi^{1/2} \mathcal{D}_a (\chi^{-1/2} A_i{}^\alpha) = 0~.
\end{align}
This implies that the $w=0$ section $\chi^{-1/2} A_i{}^\alpha$ is
covariantly constant.

We should draw attention to the fact that the hypermultiplet sector is
on-shell and so is associated with a specific Lagrangian.  The
hyperk\"ahler potential, for instance, captures all the details of a
locally supersymmetric two-derivative Lagrangian of hypermultiplets.
In closing this section we should also mention that many of the
equations obtained here can also be found in
\cite{LopesCardoso:2000qm} where the results were derived in a
slightly different context. In the next section
we will be discussing a supermultiplet that has never been subjected
to this analysis.

\section{The chiral $\mathbb{T}( \ln \bar\Phi_w)$ multiplet}
\label{sec:T-log-chiral}
\setcounter{equation}{0}
In a previous paper \cite{Butter:2013lta} a new class of
higher-derivative invariants was constructed from the so-called
kinetic multiplet.  This multiplet, denoted by
$\mathbb T(\ln\bar\Phi_w)$, is a composite chiral multiplet of weight $w=2$
constructed from the highest component of the logarithm of an
anti-chiral multiplet $\bar\Phi_w$ of arbitrary weight $w$.
In this section, we will briefly review that construction and then
analyze the conditions for a supersymmetric configuration.

Let us start by recalling that the components of a general (conformal
primary) chiral multiplet $\Phi_w$ consist of a complex scalar $A$, a
chiral fermion $\Psi_i$, a complex symmetric $\mathrm{SU}(2)$ tensor
$B_{ij}$, an anti-selfdual tensor $F_{ab}^-$, a second chiral fermion
$\Lambda_i$, and a complex scalar $C$, whose Weyl weights range from
$w$ to $w+2$.\footnote{ 
  The tensor $F_{ab}^-$, and likewise $\hat F_{ab}^-$, used in this
  section should not be confused with the (modified) field strength
  \eqref{eq:vector-field-strength} of the vector multiplet. The latter
  multiplet is related to a {\it reduced} chiral field, which implies
  that it is subject to a Bianchi identity. } 
Their supersymmetry transformation rules are
\cite{deWit:1980tn,deWit:2010za}
\begin{align}
  \label{eq:conformal-chiral}
  \delta A =&\,\bar\epsilon^i\Psi_i\,, \nonumber\\[.2ex]
  \delta \Psi_i =&\,2\,\Slash{D} A\epsilon_i + B_{ij}\,\epsilon^j +
  \tfrac12   \gamma^{ab} F_{ab}^- \,\varepsilon_{ij} \epsilon^j + 2\,w
  A\,\eta_i\,,  \nonumber\\[.2ex]
  \delta B_{ij} =&\,2\,\bar\epsilon_{(i} \Slash{D} \Psi_{j)} -2\,
  \bar\epsilon^k \Lambda_{(i} \,\varepsilon_{j)k} + 2(1-w)\,\bar\eta_{(i}
  \Psi_{j)} \,, \nonumber\\[.2ex]
  \delta F_{ab}^- =&\,\tfrac12
  \varepsilon^{ij}\,\bar\epsilon_i\Slash{D}\gamma_{ab} \Psi_j+
  \tfrac12 \bar\epsilon^i\gamma_{ab}\Lambda_i
  -\tfrac12(1+w)\,\varepsilon^{ij} \bar\eta_i\gamma_{ab} \Psi_j \,,
  \nonumber\\[.2ex]
  \delta \Lambda_i =&\,-\tfrac12\gamma^{ab}\Slash{D}F_{ab}^-
   \epsilon_i  -\Slash{D}B_{ij}\varepsilon^{jk} \epsilon_k +
  C\varepsilon_{ij}\,\epsilon^j
  +\tfrac14\big(\Slash{D}A\,\gamma^{ab}T_{abij}
  +w\,A\,\Slash{D}\gamma^{ab} T_{abij}\big)\varepsilon^{jk}\epsilon_k
  \nonumber\\*
  &\, -3\, \gamma_a\varepsilon^{jk}
  \epsilon_k\, \bar \chi_{[i} \gamma^a\Psi_{j]} -(1+w)\,B_{ij}
  \varepsilon^{jk}\,\eta_k + \tfrac12 (1-w)\,\gamma^{ab}\, F_{ab}^-
    \eta_i \,, \nonumber\\[.2ex]
    \delta C =&\,-2\,\varepsilon^{ij} \bar\epsilon_i\Slash{D}\Lambda_j
  -6\, \bar\epsilon_i\chi_j\;\varepsilon^{ik}
    \varepsilon^{jl} B_{kl}   \nonumber\\*
  &\, -\tfrac14\varepsilon^{ij}\varepsilon^{kl} \big((w-1)
  \,\bar\epsilon_i \gamma^{ab} {\Slash{D}} T_{abjk}
    \Psi_l + \bar\epsilon_i\gamma^{ab}
    T_{abjk} \Slash{D} \Psi_l \big) + 2\,w \varepsilon^{ij}
    \bar\eta_i\Lambda_j \,.
\end{align}
From these formulae, it is easy to see that if a chiral multiplet has
weight $w=0$, then requiring $\delta \Psi_i=0$ amounts to choosing $A$
to be constant and $B_{ij} = F_{ab}^- = \Lambda_i= C = 0$, as was
argued in \cite{deWit:2010za}. For chiral multiplets of non-zero
weight, the situation is more subtle, as we will soon see.

To construct $\mathbb T(\ln \bar\Phi_w)$, it is more convenient
to deal with the components of $\hat\Phi \equiv\ln\Phi_w$ rather than
with $\Phi_w$ itself. These are related in a non-linear way:
$\hat A= \ln A$, $\hat \Psi_i= A^{-1} \Psi_i$, etc.
Because $\hat A$ does not transform homogeneously under local
dilatations and U(1) transformations, the superconformal
transformations of the higher components
will be slightly modified.
The Q- and S-supersymmetry transformations of the
components $\hat A$, $\hat \Psi_i$,$\cdots$ are
\begin{align}
  \label{eq:conformal-nonlinear-chiral transf}
  \delta \hat{A} =&\,\bar\epsilon^i\hat{\Psi}_i\,, \nonumber\\[.2ex]
  \delta \hat{\Psi}_i =&\,2\,\Slash{D} \hat{A}\epsilon_i + 
  \hat{B}_{ij}\,\epsilon^j + \tfrac12   \gamma^{ab} \hat{F}_{ab}^-
  \,\varepsilon_{ij} \epsilon^j + 2\,w\,\eta_i\,,
  \nonumber\\[.2ex]
  \delta \hat{B}_{ij} =&\,2\,\bar\epsilon_{(i} \Slash{D} \hat{\Psi}_{j)}
  -2\, \bar\epsilon^k \hat{\Lambda}_{(i} \,\varepsilon_{j)k} + 
   2\,\bar\eta_{(i} \hat{\Psi}_{j)} \,,
  \nonumber\\[.2ex]
  \delta \hat{F}_{ab}^- =&\,\tfrac12
  \varepsilon^{ij}\,\bar\epsilon_i\Slash{D}\gamma_{ab} \hat{\Psi}_j+
  \tfrac12 \bar\epsilon^i\gamma_{ab}\hat{\Lambda}_i
  -\tfrac12\,\varepsilon^{ij} \bar\eta_i\gamma_{ab} \hat{\Psi}_j \,,
  \nonumber\\[.2ex]
  \delta \hat{\Lambda}_i =&\,-\tfrac12\gamma^{ab}\Slash{D}\hat{F}_{ab}^-
   \epsilon_i  -\Slash{D}\hat{B}_{ij}\varepsilon^{jk} \epsilon_k +
  \hat{C}\varepsilon_{ij}\,\epsilon^j
  +\tfrac14\big(\Slash{D}\hat{A}\,\gamma^{ab}T_{abij}
  +w\,\Slash{D}\gamma^{ab} T_{abij}\big)\varepsilon^{jk}\epsilon_k
  \nonumber\\
  &\, -3\, \gamma_a\varepsilon^{jk}
  \epsilon_k\, \bar \chi_{[i} \gamma^a\hat{\Psi}_{j]} -\,\hat{B}_{ij}
  \varepsilon^{jk}\,\eta_k + \tfrac12 \,\gamma^{ab}\, \hat{F}_{ab}^-
    \eta_i \,, \nonumber\\[.2ex]
    \delta \hat{C} =&\,-2\,\varepsilon^{ij} \bar\epsilon_i\Slash{D}\hat{\Lambda}_j
   -6\, \bar\epsilon_i\chi_j\;\varepsilon^{ik}
    \varepsilon^{jl} \hat{B}_{kl}
	+\tfrac14\varepsilon^{ij}\varepsilon^{kl} 
   \big(\bar\epsilon_i \gamma^{ab} {\Slash{D}} T_{abjk}
    \hat{\Psi}_l - \bar\epsilon_i\gamma^{ab}
    T_{abjk} \Slash{D} \hat{\Psi}_l \big)~.
\end{align}
Note in particular the transformation rule of $\hat\Psi_i$, which
transforms inhomogeneously under S-supersymmetry into a $w$-dependent constant.
For the special case of $w=0$, these components transform in
the same way as those in \eqref{eq:conformal-chiral}.

Taking the complex conjugate gives the components and transformation
rules of the anti-chiral multiplet $\ln \bar\Phi_w$.  To construct the
multiplet $\mathbb T(\ln\bar\Phi_w)$, one begins by identifying its
lowest component with the highest component of
$\ln\bar\Phi_w$. Subsequent components are defined using supersymmetry.
Here we concern ourselves only with the bosonic components and their
bosonic constituents. These are given by
\begin{align} 
  \label{eq:T-components}
  A\vert_{\mathbb{T}(\ln\bar\Phi)} &= \hat{\bar C}~, \nonumber\\[2mm] 
  B_{ij}\vert_{\mathbb{T}(\ln\bar\Phi)} &=
  - 2\,\varepsilon_{ik}\varepsilon_{jl} \big(\Box_\mathrm{c} + 3\,D\big) \hat B^{kl}
  - 2\, \hat F^+_{ab}\, R(\mathcal{V})^{ab\,k}{}_{i}\,
  \varepsilon_{jk}~, \nonumber\\[2mm]  
  F_{ab}^-\vert_{\mathbb{T}(\ln\bar\Phi)} &=
  - \big(\delta_a{}^{[c} \delta_b{}^{d]}
  - \tfrac12\varepsilon_{ab}{}^{cd}\big)
  \eol & \qquad
  \times \big[4\, D_c D^e \hat F^+_{ed}
  + (D^e \hat {\bar A}\,D_cT_{de}{}^{ij}
  + D_c \hat {\bar A} \,D^eT_{ed}{}^{ij})\varepsilon_{ij}
  - w D_c D^e T_{ed}{}^{ij} \varepsilon_{ij} \big] \nonumber\\ 
  &\quad
	+ \Box_\mathrm{c} \hat {\bar A} \,T_{ab}{}^{ij}\varepsilon_{ij}
	- R(\mathcal{V})^-{}_{\!\!ab}{}^i{}_k \,\hat B^{jk} \,\varepsilon_{ij}
	+ \tfrac1{8} T_{ab}{}^{ij} \,T_{cdij} \hat F^{+cd}~, \nonumber\\[2mm]  
C\vert_{\mathbb{T}(\ln\bar\Phi)} &=
	4 (\Box_{\rm c} + 3 D) \Box_{\rm c} \hat {\bar A} 
	+ 6 (D_a D) \, D^a \hat {\bar A}
	- 16 \,D^a \big(R(D)_{ab}^+ D^b \hat {\bar A}\big)
	\nonumber\\
	& \quad
	- D^a (T_{ab ij} T^{cbij} D_c \hat {\bar A})
	- \tfrac{1}{2} D^a(T_{ab ij} T^{cb ij}) D_c \hat {\bar A}
	+ \tfrac{1}{16} (T_{abij} \varepsilon^{ij})^2 \hat {\bar C} \nonumber\\
	& \quad
	+ \tfrac{1}{2} D_a D^a (T_{bc ij} \hat F^{bc+}) \varepsilon^{ij}
	+ 4 \, D_a \big(D^b T_{bc ij} \hat F^{ac+} + D^b
        \hat F_{bc}^+ T^{ac}{}_{ij} \big) \varepsilon^{ij}  \nonumber\\ 
          & \quad 
	- w\big[
		R(\mathcal{V})_{ab}^+{}^i{}_j R(\mathcal{V})^{ab+}{}^j{}_i
		+ 8  R(D)^+_{ab} R(D)^{ab+}\big] \nonumber\\
		&\quad -w\big[ D^a T_{ab ij} D_c T^{cb ij}
		+  D^a (T_{abij} D_c T^{cb ij})\big] ~.
\end{align}

Following the same strategy as before,
let us analyze the conditions for a supersymmetric
configuration. Requiring $\delta \hat\Psi_i = 0$ leads to
\begin{align}
  \label{eq:lnPhieta}
  \hat\eta_i =- \frac1{w}\Big[\Slash{D} \hat A\epsilon_i
	+ \tfrac1{2} \hat B_{ij}\epsilon^j
	+ \tfrac1{4}\gamma^{ab} \hat F_{ab}^- \varepsilon_{ij}
        \epsilon^j\Big] ~.
\end{align}
Next we sequentially impose $\delta \hat \Lambda_i = 0$,
$\delta\chi^i = \delta R(Q)_{ab}{}^i = 0$ and finally
$\delta (D_a \hat \Psi_i) = 0$ using this choice for $\hat\eta_i$.
We find several algebraic conditions,
\begin{alignat}{2}
\hat B_{ij} \hat F_{ab}^- &= \hat B_{ij}T_{ab}{}^{kl} = 0~,&\qquad\,
\hat C&=-\tfrac{1}{2w} \hat F_{ab}^-\,\hat F ^{ab\,-}
	-\tfrac1{4w}\hat B_{kl}\hat B_{mn}\varepsilon^{kn}\varepsilon^{lm}~, \eol
\hat F_{a[b}^- T_{c]}{}^{a\,ij} &= 0~, &\qquad 
D &= \tfrac{1}{24w} \hat F^{ab -} T_{ab}{}^{ij} \veps_{ij}~,
\label{eq:lnPhiCon1}
\end{alignat}
in addition to the first-order differential equations
\begin{align}\label{eq:lnPhiCon2}
\mathcal{D}_\mu \hat B_{ij} - \tfrac1{w} \mathcal{D}_\mu \hat A\, \hat B_{ij} &= 0~,\eol
\mathcal{D}_a T^{bc}{}^{ij}
	- \tfrac{1}{w}  \mathcal{D}_a \hat A\, T^{bc}{}^{ij}
	+ \tfrac{2}{w} \mathcal{D}^{[b} \hat A \,T^{c]}{}_a{}^{ij} 
	- \tfrac{2}{w} \mathcal{D}_d \hat A \,T^{d[b}{}^{ij} \delta^{c]}{}_a &= 0~, \eol
\mathcal{D}_{a} \hat F^{bc-}
	- \tfrac1{w}\mathcal{D}_a \hat {\bar{A}} \,\hat F^{bc-} 
	+ \tfrac2{w} \mathcal{D}^{[b}\hat A\, \hat F^{c]}{}_a{}^-
	- \tfrac2{w} \mathcal{D}_d \hat A \,\hat F^{-d[b} \delta^{c]}{}_a
	&= 0~,
\end{align}
and the second-order differential equation
\begin{gather}
\mathcal{D}_a \mathcal{D}_b \hat A + w \,e_a{}^\mu f_{\mu b} - \tfrac1{w}\mathcal{D}_a \hat A \mathcal{D}_b \hat A + \tfrac1{2w} \mathcal{D}_c \hat A \mathcal{D}^c \hat A\,\eta_{ab}
	+\tfrac{3}{4} \,w\,D \,\eta_{ab}-\tfrac{1}{2w} \hat F^-_{ac}\,\hat F^{+\,c}{}_b = 0~.
\label{eq:lnPhiCon3}
\end{gather}
One additional condition is also found:
\begin{align}\label{eq:lnPhiCon4}
\mathcal{D}^c (\hat A - \hat {\bar A}) \,\hat F_{cb}^- =
	-\tfrac{1}{4} \,w\,\mathcal{D}^c (\hat A - \hat {\bar A}) \,T_{cb\,ij}\,\veps^{ij}~.
\end{align}

From \eqref{eq:lnPhiCon1}, we deduce that
\begin{align}\label{eq:lnPhiComp}
\hat B_{ij} = 0~, \qquad
\hat F_{ab}^- = \frac{24 \,w \,D\,T_{ab}{}^{ij} \veps_{ij}}{(T_{cd}{}^{kl} \veps_{kl})^2}~, \qquad
\hat C = - \frac{288 \,w \,D^2}{(T_{ab}{}^{ij} \veps_{ij})^2}~.
\end{align}
Multiplying the second equation of \eqref{eq:lnPhiCon2}
by $T_{bc}{}^{kl}$ leads to
$\mathcal{D}_a\big[\hat A - \tfrac{1}{2} w \ln(T^{bcij}\varepsilon_{ij})^2 \big] = 0$.
Because $\hat A- \tfrac{1}{2} w \ln(T^{bcij}\varepsilon_{ij})^2$
is inert under dilatations and U(1) rotations, one recovers
\begin{align}\label{eq:lnPhiA}
\mathcal{D}_a\big[\hat A - \tfrac{1}{2} w \ln(T^{bcij}\varepsilon_{ij})^2 \big] = 0
\quad\implies\quad
\hat A = \tfrac{1}{2} w \ln (T_{ab}{}^{ij} \veps_{ij})^2 + \textrm{const}~.
\end{align}
With these choices, the equations \eqref{eq:lnPhiCon1}--\eqref{eq:lnPhiCon4}
are identically satisfied, once we use the conditions established for
the Weyl multiplet in section \ref{sec:vector-supermultiplet}.
At this point we should remark that we could have immediately derived
these results by noting that
\begin{align}
\hat A - \tfrac{1}{2} w \ln (T_{ab}{}^{ij} \veps_{ij})^2 = 
	\ln \left(\frac{A}{((T_{ab}{}^{ij} \veps_{ij})^2)^{w/2}}\right)
\end{align}
is the lowest component of a $w=0$ chiral multiplet and therefore must be
a constant. The higher components of this new $w=0$ multiplet must vanish,
which leads after some algebra to the relations \eqref{eq:lnPhiComp}.

Now we are in a position to evaluate the supersymmetric configuration of
$\mathbb T(\ln\bar\Phi_w)$.
From \eqref{eq:lnPhiComp} one finds that the lowest component of the
kinetic multiplet is completely determined to be
\begin{align}\label{eq:TlnA}
A \vert_{\mathbb T(\ln\bar\Phi_w)} 
	= - \frac{288 \,w \,D^2}{(T_{ab}{}_{ij} \veps^{ij})^2}~.
\end{align}
The remainder of the components of $\mathbb T(\ln\bar\Phi_w)$
can be found by explicit use of the formulae \eqref{eq:T-components},
but it is much simpler to note that since $\mathbb T(\ln\bar\Phi_w)$ is a
$w=2$ chiral multiplet, it must be proportional to the square
of the Weyl multiplet, schematically denoted $W^2$, whose lowest component is
$(T_{ab}{}^{ij} \veps_{ij})^2$. For example, we can relate the component
$B_{ij}$ of $\mathbb T(\ln\bar\Phi_w)$ to the same component of $W^2$,
\begin{align}
B_{ij}\vert_{\mathbb T(\ln\bar\Phi_w)} = B_{ij}\vert_{W^2} \times
	\frac{A\vert_{\mathbb T(\ln\bar\Phi_w)}}{(T_{cd}{}^{kl} \veps_{kl})^2} = 0~.
\end{align}
In the last equality we have used the fact that in the supersymmetric
configuration $B_{ij}\vert_{W^2}$ is proportional to
$\veps_{ik} R(\cV)_{ab}{}^k{}_j$, which vanishes.
In a similar way, one finds
\begin{align}
F_{ab}^- \vert_{\mathbb T(\ln\bar\Phi_w)} = 48 D\, T_{ab}{}^{ij} \veps_{ij} \,
	\frac{A\vert_{\mathbb T(\ln\bar\Phi_w)}}{(T_{cd}{}^{kl} \veps_{kl})^2}~, \quad
C\vert_{\mathbb T(\ln\bar\Phi_w)} = 576 D^2 \,\frac{A\vert_{\mathbb T(\ln\bar\Phi_w)}}{(T_{cd}{}^{kl} \veps_{kl})^2}~.
\end{align}
Note that these higher components are completely determined by
the lowest component $A\vert_{\mathbb T(\ln \bar\Phi_w)}$, given in
\eqref{eq:TlnA}. Two special cases are worthy
of note. If $\Phi_w$ is actually a weight $w=0$ multiplet,
then $\mathbb T(\ln \bar\Phi_w)$ vanishes completely, as was noted
in \cite{deWit:2010za}.
Similarly, if we apply the conditions of section \ref{sec:short-mutiplets}
(equivalently, the conditions of \cite{LopesCardoso:2000qm}),
then $D=0$ causes the entire kinetic multiplet to vanish for any 
value of the Weyl weight. This will be a crucial point for the
non-renormalization theorem presented in the next section.

\section{A new non-renormalization theorem}
\label{sec:non-renormalization}
\setcounter{equation}{0}
The preceding sections have mainly been concerned with deriving the
conditions of off-shell $N=2$ supersymmetry for various multiplets
independently of any action. We devoted particular attention to
the chiral multiplet $\mathbb T(\ln \bar\Phi_w)$, which has been constructed
only recently. This multiplet leads to a
new class of $4D$ higher-derivative invariants. Our goal in this section
is to establish a non-renormalization theorem: in a fully
supersymmetric configuration, these higher-derivative invariants
always vanish, as do their first derivative with respect to any field
or coupling constant. To accomplish this, we will make one assumption.
In addition to the apparent field content -- a non-vanishing
chiral multiplet $\Phi_w$ coupled to conformal supergravity --
we require at least one
multiplet of the set discussed in section \ref{sec:short-mutiplets}.
The motivation for this last requirement is physical.
A Poincar\'e supergravity action requires both a
vector multiplet and at least one other short multiplet.
So even if such a multiplet is not present in the specific
higher-derivative terms under discussion, it must be present in the 
sector of the action responsible for generating Poincar\'e supergravity.
This means that it too must take its supersymmetric value.
Making this assumption means that the restrictive conditions
discussed in section \ref{sec:short-mutiplets} apply.
In particular, we will require that $D=0$.

It will be convenient to exploit superfield and superspace terminology
as discussed in \cite{Butter:2013lta}.
Superspace actions generically
fall into two classes: they can be integrals over chiral
superspace or integrals over the full superspace. Schematically,
we can write a chiral superspace action up to a normalization factor
as
\begin{align}\label{eq:ChiralInt}
\int \rd^4x\, \rd^4 \q\, \cE\, F
\end{align}
where $F$ is some quantity built out of chiral multiplets
(fundamental or composite) and $\cE$ is the chiral superspace
measure. The other option is a full superspace integral
\begin{align}\label{eq:FullSuperInt}
\int \rd^4x\, \rd^4 \q\, \rd^4\bar\q\, E\, \cH~,
\end{align}
where $\cH$ is real and $E$ is the full superspace measure.
In order to satisfy the requirements
of superconformal invariance, $F$ must have Weyl weight $w=2$
and $\cH$ must have Weyl weight $w=0$. In addition, both
$F$ and $\cH$ must be annihilated by S-supersymmetry.

The distinction between these two types of invariants is not
a sharp one. Any full superspace integral can be recast as a
chiral one by making use of the so-called $N=2$ kinetic operator
$\mathbb T$,
normalized here so that\footnote{The kinetic operator
defined in \cite{deWit:2010za} acts on an anti-chiral multiplet of
weight $w=0$. It can be extended to
act on any conformal primary (chiral or not) with $w=-c$
to yield a new chiral multiplet of weight $w+2$.
This is equivalent to the chiral projection operator
defined in superspace \cite{Muller:1988ux, KT-M:DReps}.}
\begin{align}
\int \rd^4x\, \rd^4 \q\, \rd^4\bar\q\, E\, \cH
	= -\frac{1}{2} \int \rd^4x\, \rd^4 \q\, \cE\, \mathbb T(\cH)~.
\end{align}
Therefore, when we discuss chiral superspace invariants,
we usually mean ones which \emph{cannot} be converted back
into full superspace invariants by removing a kinetic operator.
It will be convenient to call such chiral multiplets
\emph{intrinsically chiral}.

A common example of intrinsically chiral integrands are of the form
$F(X, A\vert_{W^2})$ where $X^I$ are vector multiplets and
$A\vert_{W^2} = (T_{ab}{}^{ij} \veps_{ij})^2$ is the lowest component
of the square of the Weyl multiplet.  This class $F(X, A\vert_{W^2})$
is actually quite important: it was shown in
\cite{LopesCardoso:1998wt,LopesCardoso:1999ur} to accurately
  describe the subleading corrections to the Wald entropy in the limit
  of large charges required for matching the degeneracy of the
  microscopic string and brane states. This precise matching was in
  retrospect quite surprising since there are in principle a number of
  higher-derivative actions that do not fall into this class. In fact,
  this was the motivation in \cite{deWit:2010za} where a
  non-renormalization theorem established that a large class of full
  superspace integrals \eqref{eq:FullSuperInt} do not contribute to
  the Wald entropy.

It is now important to address what other intrinsically chiral
invariants might exist and whether they might possess non-renormalization
theorems as well. As discussed in \cite{Butter:2013lta},
the kinetic multiplet $\mathbb T(\ln \bar\Phi_w)$
is actually a new contribution to intrinsically chiral functions
$F$. To see why, we note that the naive equality
\begin{align}
-\frac{1}{2} \int \rd^4x\, \rd^4\q\, \cE\, \Phi'\, \mathbb T(\ln\bar\Phi_w)
	\stackrel{?}{=} \int \rd^4x\, \rd^4\q\, \rd^4\bar\q\, E\, \Phi'\, \ln\bar\Phi_w
\end{align}
(where $\Phi'$ is some $w=0$ chiral multiplet)
does not hold since the integrand on the right-hand side
is not actually weight zero due to the inhomogeneous dilatation
transformation of $\ln\bar\Phi_w$. This means that the
left-hand side is actually an intrinsically chiral quantity.

It would seem that this observation might open the door for many new
intrinsically chiral contributions, but it turns out this is
not the case. The reason is that any two such multiplets are actually
related to each other by the kinetic operator of a weight-zero multiplet.
Taking $\Phi_w'$ and $\Phi_w$ to be chiral multiplets of the same nonzero
weight (for simplicity), the difference 
\begin{align}
\mathbb T(\ln\bar\Phi'_w) - \mathbb T(\ln\bar\Phi_w) =
	\mathbb T(\ln (\bar\Phi'_w / \bar\Phi_w))
\end{align}
is actually the kinetic multiplet of a weight-zero
multiplet. This permits, for example, manipulations like
\begin{align}
\int \rd^4x\, \rd^4\q\, \cE~ \Phi'~ \mathbb T(\ln\bar\Phi'_w) 
	&= \int \rd^4x\, \rd^4\q\, \cE~ \Phi' ~\mathbb T(\ln\bar\Phi_w) 
		- 2 \int \rd^4x\, \rd^4\q\, \rd^4\bar\q\, E ~\Phi' ~ \ln(\bar\Phi'_w/\bar\Phi_w)~,
\end{align}
where $\Phi'$ is a $w=0$ chiral multiplet.
This allows any operators $\mathbb T(\ln\bar\Phi'_w)$ to be traded
for one universal choice $\mathbb T(\ln\bar\Phi_w)$ and the rest lifted
to full superspace integrals, where the non-renormalization theorem of
\cite{deWit:2010za} applies.

We will now establish a new non-renormalization theorem:
the contribution of $\mathbb T(\ln\bar \Phi_w)$
to any chiral integral \eqref{eq:ChiralInt} always vanishes
as does the first derivative with respect to any field or coupling
constant. Using the condition $D=0$ found in section \ref{sec:short-mutiplets},
we find that \emph{the entire kinetic multiplet $\mathbb T(\ln\bar\Phi_w)$ vanishes
in a supersymmetric vacuum.} In other words, in a supersymmetric vacuum,
we can replace
\begin{align}
F(\Phi, \mathbb T(\ln \bar \Phi_w)) \quad \longrightarrow \quad
	F(\Phi, 0)
\end{align}
in any chiral superspace integral \eqref{eq:ChiralInt}.
We still must be careful to analyze what happens under \emph{variations}
of the fields in a supersymmetric configuration. For simplicity, 
we consider first the case
\begin{align}\label{eq:NLKsuper}
-2 \int \rd^4x\, \rd^4\q\, \cE~ \Phi'~ \mathbb T(\ln \bar \Phi_w)
\end{align}
with a weight-zero chiral multiplet $\Phi'$ whose component action
was constructed in \cite{Butter:2013lta}. (An overall factor of
$-2$ is necessary to match the component action normalization of
\cite{Butter:2013lta}.)
In principle, there are three ways in which this quantity could be
varied: we may vary either of the two multiplets $\Phi'$
and $\bar\Phi_w$ explicit in the expression, or we may vary
the supergravity fields which are implicit.
Variations of $\Phi'$ clearly give zero since
$\mathbb T(\ln\bar\Phi_w)$ vanishes in the supersymmetric
background. Variations of $\bar\Phi_w$ within the kinetic
multiplet also give zero. This can be seen by parametrizing
the variation as
$\delta \bar\Phi_w = \bar\Phi_w \bar\L$ where $\bar\Lambda$
is a $w=0$ anti-chiral multiplet. This leads to
$\mathbb T (\delta \ln\bar\Phi_w) = \mathbb T (\bar \Lambda)$
and so we can write
\begin{align}
\delta_{\Phi_w} \int \rd^4x\, \rd^4\q\, \cE ~ \Phi' ~\mathbb T(\ln \bar \Phi_w)
	&= \int \rd^4x\, \rd^4\q\, \cE ~\Phi' ~\mathbb T(\bar \Lambda)
	= \int \rd^4x\, \rd^4\bar\q\, \bar\cE ~\bar{\mathbb T}(\Phi') ~\bar \Lambda
\end{align}
where we ``integrate by parts'' the kinetic operator as in
\cite{deWit:2010za}. Since $\Phi'$ has zero Weyl weight, its
supersymmetric value is a constant and so $\bar{\mathbb T}(\Phi') = 0$.
The last possibility is to vary the components of the Weyl multiplet
itself, with $\Phi'$ fixed at
its supersymmetric value. Taking the result for the component
action of \eqref{eq:NLKsuper} given in \cite{Butter:2013lta} and imposing
the supersymmetry conditions on the components of $\Phi'$, one finds
\begin{align}\label{eq:NLK1}
  e^{-1} \mathcal{L}_{} =&\,
	w\,A' \,\Big(\tfrac{2}{3} \mathcal{R}^2 - 2\, \mathcal{R}^{ba} \mathcal{R}_{ab} - 6\, D^2
		+ 2 \, R(A)^{ab} R(A)_{ab} -  R(\mathcal{V})^{+ab}{}^i{}_j\, R(\mathcal{V})^+_{ab}{}^j{}_i
		\nonumber \\
                & \quad \qquad\qquad
		+ \tfrac{1}{128}  T^{ab ij} T_{ab}{}^{kl} T^{cd}{}_{ij} T_{cd kl}
		+  T^{ac ij} \mathcal{D}_a \mathcal{D}^b T_{bc ij}
		-  T^{ac ij} f_a{}^b T_{bc ij}
		\Big) \,~,
\end{align}
where $A'$ must be a constant. Note already that the terms
$D^2$, $(R(A)_{ab})^2$ and $(R(\mathcal{V})^+_{ab}{}^i{}_j)^2$ are quadratic in
quantities which vanish in the supersymmetric background,
and so any variation of these quantities must vanish. It turns out that
the same holds for the remaining terms.
The Lagrangian \eqref{eq:NLK1} can be written as
\begin{align}\label{eq:NLK2}
  e^{-1} \mathcal{L}_{} =&\,
	w\,A' \,\Big(
		2 (Z_{ab} \eta^{ab})^2 - 2 Z^{ba} Z_{ab} 
		- \tfrac{1}{2} Z^1_a Z^{2a}
		- 6\, D^2
		\eol & \qquad
		+ 2 \, R(A)^{ab} R(A)_{ab} -  R(\mathcal{V})^{+ab}{}^i{}_j\, R(\mathcal{V})^+_{ab}{}^j{}_i
		+ \mathcal{D}^a \mathcal{O}_a \Big)
\end{align}
where the three complex quantities
\begin{align}
Z_{ab} &=\cR_{ab} - \tfrac{1}{6} \eta_{ab} \cR 
	+ \tfrac{1}{8} T_{ac\,ij} T_{b}{}^c{}^{ij}
	+ 2 w^{-1} \mathcal{D}_a \mathcal{D}_b \hat {\bar A}
	- 2 w^{-2} \mathcal{D}_a \hat{\bar A} \mathcal{D}_b \hat{\bar A}
	+ w^{-2} \eta_{ab} (\mathcal{D}_c \hat{\bar A})^2~, \eol
Z_a^1 &= \mathcal{D}^b T_{ba\,ij} \,\veps^{ij} + w^{-1} \mathcal{D}^b \hat{\bar A}\, T_{ba\,ij} \,\veps^{ij} ~, \eol
Z_a^2 &= \mathcal{D}^b T_{ba}{}^{ij} \,\veps_{ij} + w^{-1} \mathcal{D}^b \hat{\bar A}\, T_{ba}{}^{ij} \,\veps_{ij} ~,
\end{align}
vanish in a supersymmetric configuration, 
using the supersymmetry conditions
\eqref{eq:lnPhiCon1} -- \eqref{eq:lnPhiCon4},
along with the additional condition $D=0$
(which implies $\mathcal{D}_a \hat A = \mathcal{D}_a \hat {\bar A}$).
The last term of \eqref{eq:NLK2}, which involves $\cD_a \cO^a$ for
\begin{align}
\mathcal{O}_a &=
	T_{ac}{}^{ij} \mathcal{D}_b T^{bc}{}_{ij}
	+ w^{-1} T_{ac\,ij} T^{bc\,ij} \,\mathcal{D}_b \hat {\bar A}
	- 4 w^{-1} \cR \,\mathcal{D}_a \hat{\bar A}
	+ 8 w^{-1} \cR_{ba} \mathcal{D}^b \hat{\bar A}
	\eol & \quad
	- 8 w^{-2} \mathcal{D}_a \hat{\bar A} \mathcal{D}^2 \hat{\bar A}
	+ 8 w^{-2} \mathcal{D}^b \hat{\bar A} \,\mathcal{D}_b \mathcal{D}_a \hat{\bar A}
	- 8 w^{-3} \mathcal{D}_a \hat{\bar A} \,(\mathcal{D}_c \hat{\bar A})^2~,
\end{align}
gives a total derivative because $A'$ is constant.
The remaining pieces are each quadratic in terms that vanish in
the supersymmetric vacuum, so their variation with respect to any
of the supergravity fields must vanish.

We have now established a non-renormalization theorem for
the expression \eqref{eq:NLKsuper}. This is straightforwardly
extended to the more general class of functions
\begin{align}\label{eq:NLKGenCase}
\int \rd^4x\, \rd^4\q\, \cE\, F(\Phi^I, \mathbb T(\ln \bar \Phi_w))~.
\end{align}
Here the superfields $\Phi^I$ are a set of chiral superfields
which may possess any weight. For instance, they may
consist of vector multiplets $X^I$ and the chiral supergravity invariant
$W^{\alpha\beta} W_{\alpha\beta}$. We have already observed that
in a supersymmetric vacuum $\mathbb T(\ln\bar\Phi_w)$ vanishes.
In this context, the functions $F$ should be 
analytic at $\mathbb T(\ln\bar\Phi_w)=0$. Therefore, we may
construct a series
expansion, a characteristic term of which would be
\begin{align}
\int \rd^4x\, \rd^4\q\, \cE ~ \Phi_{2-2n} ~\big[\mathbb T(\ln\bar\Phi_w)\big]^n~.
\end{align}
But any such term can always be written as \eqref{eq:NLKsuper}
for the choice $\Phi' \propto \Phi_{2-2n} \big[\mathbb T(\ln\bar\Phi_w)\big]^{n-1}$.
Since our treatment of \eqref{eq:NLKsuper} holds for arbitrary
$\Phi'$, the non-renormalization theorem applies to this term
and therefore to the broad class \eqref{eq:NLKGenCase}.

\section{Dimensional reduction of the $5D$ mixed gauge-gravitational CS invariant}
\label{sec:5d4d}
\setcounter{equation}{0}
The kinetic multiplet $\mathbb T(\ln \bar\Phi_w)$ discussed in the preceding
sections plays a natural role in extending the known classes of chiral superspace
higher-derivative invariants. As alluded to in the introduction and discussed
briefly in \cite{Butter:2013lta}, evidence for the existence of a new class
of higher-derivative invariants
was actually seen in \cite{Banerjee:2011ts} where the dimensional
reduction of the supersymmetric version of the $5D$ Chern-Simons action
$\Tr ( W \wedge R \wedge R)$ was considered.
The authors of \cite{Banerjee:2011ts} identified three distinct types of
terms in the dimensional reduction: one corresponded to a usual
chiral superspace integral of a holomorphic prepotential $F(X, A\vert_{W^2})$,
another was identified as a full superspace integral $\cH(X, \bar X)$,
and a third remained a mystery. As discussed in \cite{Butter:2013lta},
this identification was actually incorrect: the second and third invariants
described in \cite{Banerjee:2011ts} are actually part of a single
irreducible chiral invariant constructed from a kinetic
multiplet $\mathbb T(\ln\bar\Phi_w)$. Our goal in this section is to
back up this claim by keeping a much wider range of terms in the dimensional
reduction and checking against the proposed $4D$ action.

The supersymmetric version of the $5D$ Chern-Simons action $\Tr(W \wedge R \wedge R)$,
constructed originally in \cite{Hanaki:2006pj}, is given in
the conventions of \cite{deWit:2009de} by
\begin{align}
  \label{eq:5Dvww}
   E^{-1}\, \mathcal{L}_\mathrm{vww} &=
  \tfrac14 c_I  Y_{ij}{}^I \,  T^{AB}  R_{ABk}{}^j(V) \,\varepsilon^{ki}
  \nonumber \\[.5ex]
&\quad
  + c_I \sigma^I\Big[ 
     \tfrac1{64}  R_{AB}{}^{CD}(M)\, R_{CD}{}^{AB}(M)+\tfrac1{96}
   R_{AB j}{}^i(V) \, R^{AB}{}_i{}^j(V) \Big]  \nonumber\\[.5ex]
  &\quad
   -\tfrac1{128}\mathrm{i} \, E^{-1}
  \,\varepsilon^{MNPQR}\,c_I  W_M{}^I\left[  
   R_{NP}{}^{AB}(M)\, R_{QR AB}(M)+ \tfrac13
   R_{NP j}{}^i(V) \, R_{QR i}{}^j(V)\right] \Big]
  \nonumber\\[.5ex] 
  &\quad  + \tfrac3{16} c_I\big(10\, \sigma^I \, T_{AB} -  F_{AB}{}^I\big)\,
   R(M)_{CD}{}^{AB}\, T^{CD} \nonumber\\[.5ex]
  &\quad
  + c_I \sigma^I\Big[ 3\, T^{AB} {\mathcal{D}}^C  {\mathcal{D}}_A  T_{BC} -\tfrac32
  \big( {\mathcal{D}}_A  T_{BC}\big)^2 
  + \tfrac32  {\mathcal{D}}_C  T_{AB}\, {\mathcal{D}}^A  T^{CB}\Big]  
   \nonumber\\[.5ex]
   &\quad
   + c_I  \sigma^I\Big[
   \tfrac83  D^2 + 8\,  T^2\, D - \tfrac{33}8 ( T^2)^2 + \tfrac{81}2
   ( T^{AC}  T_{BC})^2
   +  {\mathcal{R}}_{AB}( T^{AC}  T^B{}_C - \tfrac12  \eta^{AB}  T^2) \Big]    \nonumber\\[.5ex]
  &\quad
  +  \tfrac3{4}\mathrm{i} \,\varepsilon^{ABCDE}\Big[c_I  F_{AB}{}^I
  \big( T_{CF}  {\mathcal{D}}^F  T_{DE} +\tfrac32  T_{CF}
   {\mathcal{D}}_D  T_E{}^F\big)
   -  3\,c_I  \sigma^I  T_{AB}  T_{CD}\, {\mathcal{D}}^F  T_{FE}\Big]
   \nonumber\\[.5ex]
  &\quad
   - c_I  F_{AB}{}^I\Big[ T^{AB}\, D +\tfrac{3}{8} T^{AB} \, T^2 - \tfrac9{2} \, 
   T^{AC}  T_{CD}  T^{DB} \Big]~,
\end{align}
with $E = \det(E_M{}^A)$, the determinant of the $5D$ vielbein.
The fields $\sigma^I$, $W_M{}^I$, and $Y_{ij}{}^{I}$ are the bosonic
components of a $5D$ vector multiplet, with field strength
$F_{MN}{}^{I} = 2 \pa_{[M} W_{N]}{}^I$. The index $I$ enumerates
a number of such multiplets.
The fields $T_{AB}$ and $D$ are the covariant bosonic fields of the
$5D$ Weyl multiplet. The $5D$ Lorentz and SU(2) curvature tensors
are given respectively by $R(M)_{MN}{}^{AB}$
and $R(V)_{MN}{}_i{}^j$.

We will show that the full $4D$ invariant that matches the reduction
of \eqref{eq:5Dvww} is given by
\begin{align}\label{eq:4Dvww_super}
S_{\rm vww} &= \frac{\ri}{64} \int \rd^4x\, \rd^4\q\, \cE\, 
	\,c_I \,\frac{X^I}{X^0}
	\Big(W^{\alpha\beta} W_{\alpha\beta}
		- \tfrac{1}{3} \mathbb{T}(\ln\bar X^0)\Big) + \HC
\end{align}
This corresponds to a chiral superspace action where
the holomorphic function $F$ is, in the usual normalization convention,
given by
\begin{align}\label{eq:Fvww}
F &= -\frac{1}{64} 
	\,\frac{c_I X^I}{X^0}
	\Big(\tfrac{1}{32} (T_{ab}{}^{ij} \veps_{ij})^2
		- \tfrac{1}{3} A\vert_{\mathbb{T}(\ln\bar X^0)}\Big)~.
\end{align}
This expression involves three types of fields: the ``matter'' vector
multiplets $X^I$, the Kaluza-Klein vector multiplet $X^0$, and the $4D$
Weyl multiplet superfield $W_{\alpha\beta}$ whose lowest component is
$T_{ab}{}^{ij} \veps_{ij}$.
The expression within parentheses in \eqref{eq:4Dvww_super} is composed
of two chiral invariants. The first involves the square of the
Weyl multiplet, and the second involves the
kinetic multiplet $\mathbb T(\ln\bar X^0)$.

Before proceeding to details of the actual computation, some elucidating
comments are necessary about how to organize the Lagrangian.
While \eqref{eq:5Dvww} is fairly complicated, we draw attention to
one important feature: every term is linear in a component of
the $5D$ vector multiplet.
Upon dimensional reduction we must retain this feature,
so the $4D$ Lagrangian should take the form
\begin{align}\label{eq:4Dvww_gen}
e^{-1} \cL\vert_{\rm 4D} =
	- \tfrac{1}{2} c_I Y^{ij\, I}\, L_{ij}
	- \tfrac{1}{2} \ri \,c_I F_{\mu\nu}{}^I \,\tilde E^{\mu\nu}
	+ c_I X^I \, G + c_I \bar X^I\, \bar G
\end{align}
for some composite functions $L_{ij}$,
$\tilde E_{\mu\nu} \equiv \tfrac{1}{2} \veps^{\mu\nu\rho\sigma} E^{\rho\sigma}$,
$G$ and $\bar G$.
It is natural to write the coefficient of $F_{\mu\nu}{}^I$ as
the dual of a two-form $E_{\mu\nu}$ since the Bianchi identity
on $F_{\mu\nu}{}^I$ implies that
$E_{\mu\nu}$ can be defined only up to a gauge transformation,
$E_{\mu\nu} \rightarrow E_{\mu\nu} + 2 \pa_{[\mu} \L_{\nu]}$.

We have chosen the normalizations of the composite functions in
\eqref{eq:4Dvww_gen} in a very particular way.  Supersymmetry dictates
that the functions $L_{ij}$, $E_{\mu\nu}$, $G$, and $\bar G$, must
correspond to the bosonic components of a (composite) tensor
multiplet.  This has some deep implications when one compares two
expressions of the form \eqref{eq:4Dvww_gen}, such as those we plan to
derive from \eqref{eq:5Dvww} and \eqref{eq:4Dvww_super}.  In
particular, to show full equivalence between them, we must only prove
that the two expressions for $L_{ij}$ are the same: as these are the
lowest components of some (composite) tensor multiplet, the equality
of the remaining pieces follows by supersymmetry.

Unfortunately, we cannot fully exploit this observation.
A strict proof along these lines requires that the fermionic
bilinears of $L_{ij}$ be compared as well, and in the calculation
of the Lagrangian \eqref{eq:5Dvww} these would need to be
restored. We will instead demonstrate a proof of equivalence between all bosonic
terms of $L_{ij}$, as well as some characteristic bosonic terms
of $E_{\mu\nu}$ and $G$. This establishes beyond any
doubt the equivalence between \eqref{eq:4Dvww_super} and
the reduction of \eqref{eq:5Dvww}.

We begin by reviewing some key results of the off-shell
dimensional reduction formulated in \cite{Banerjee:2011ts}. In order
to avoid confusion between $4D$ and $5D$ fields, we henceforth
will place a diacritic on all $5D$ quantities
(e.g. $E_M{}^A \rightarrow \wt E_M{}^A$).
All bosonic components of the $5D$ Weyl multiplet,
($\wt E_M{}^A$, $\wt b_M$, $\wt V_M{}_i{}^j$, $\wt T_{AB}$, and $\wt D$),
must reduce to expressions involving the $4D$ Weyl multiplet 
and a Kaluza-Klein vector multiplet $X^0$.
Below we provide a dictionary relating the $5D$ and $4D$ components.
To avoid potential confusion the index $5$ will
refer \emph{only} to the fifth component of the tangent space
index $A$ and \emph{never} to the fifth coordinate.

The fundamental bosonic fields of the Weyl multiplet are given by
\begin{align}
\wt E_M{}^A &=
\begin{pmatrix}
e_\mu{}^a & \tfrac{1}{2} W_\mu{}^0 \,|X^0|^{-1} \\[2mm]
0 & \tfrac{1}{2} \,|X^0|^{-1}
\end{pmatrix}~, \quad
\wt b_M =
\begin{pmatrix}
b_m \\[2mm]
0
\end{pmatrix}~, \nonumber\\[1ex]
\wt V_a{}_i{}^j &= \cV_a{}^j{}_i~, \quad
\wt V_5{}_i{}^j = -\frac{1}{2} \veps_{ik} Y^{kj\, 0} |X^0|^{-1}~, \eol [1ex]
\wt  T_{ab} &= -\tfrac{1}{24} \ri\, \Big(
	\veps_{ij} T_{ab}{}^{ij} \bar X^0 - F_{ab}^-{}^0
	\Big)|X^0|^{-1} + \HC~, \qquad
\wt T_{a5} = \tfrac{1}{12} \ri\,\mathcal{D}_a \ln (X^0 / \bar X^0)~, \nonumber\\[1ex]
\wt D &= \tfrac{1}{4} D - \tfrac{1}{16} |X^0|^{-1} (\mathcal{D}^a \mathcal{D}_a + \tfrac{1}{6} \cR) |X^0|
	- \tfrac{3}{512} |X^0|^{-2} F_{ab}{}^0 F^{ab\, 0}
	\eol & \quad
	+ \tfrac{1}{64} |X^0|^{-2} Y^{ij\, 0} Y_{ij}{}^0
	- \tfrac{3}{8} \wt T^{ab} \wt T_{ab}
	- \tfrac{3}{4} \wt T^{a5} \wt T_{a5}~.
\end{align}
Some derived quantities are also useful. The $5D$ spin connection and Riemann tensor
can be found in \cite{Banerjee:2011ts}, while the $5D$ SU(2) curvature tensor is given by
\begin{align}\label{eq:RV5to4}
\wt R(V)_{ab\, i}{}^j &= R(\cV)_{ab}{}^j{}_i - \tfrac{1}{4} \veps_{ik} Y^{kj\, 0}\, F_{ab}{}^0\, |X^0|^{-2}~, \eol
\wt R(V)_{a5\, i}{}^j &= -\tfrac{1}{2} \veps_{ik} |X^0| \,\mathcal{D}_a \Big(Y^{kj\,0} / |X^0|^2\Big)~.
\end{align}
The decomposition of the $5D$ vector multiplet is given by
\begin{align}
\wt \sigma^I &= -\ri \,|X^0|\, \Big(\frac{X^I}{X^0} - \frac{\bar X^I}{\bar X^0}\Big)~,
&
\wt Y^{ij\,I} &= -\tfrac{1}{2} Y^{ij\,I} + \tfrac{1}{4} \Big(\frac{X^I}{X^0}+\frac{\bar X^I}{\bar X^0}\Big) Y^{ij\, 0}~,\eol
\wt W_a{}^I &= W_a{}^I~,
&
\wt W_5{}^I &= -|X^0| \Big(\frac{X^I}{X^0} + \frac{\bar X^I}{\bar X^0}\Big)~, \eol
\wt F_{ab}{}^I &= F_{ab}{}^I -\tfrac{1}{2}  F_{ab}{}^0 \, \Big(\frac{X^I}{X^0} + \frac{\bar X^I}{\bar X^0}\Big)~,
&
\wt F_{a5}{}^I &= -|X^0| \,\mathcal{D}_a \Big(\frac{X^I}{X^0} + \frac{\bar X^I}{\bar X^0}\Big)~.
\end{align}
It is important to note that all of these equations are invariant under
the $4D$ U(1) R-symmetry group. This is because there is no U(1) factor in
the $5D$ R-symmetry group; it emerges from the dimensional reduction.

Let us now analyze the first term $L_{ij}$ of the $4D$ Lagrangian 
\eqref{eq:4Dvww_gen}. This arises only from the first term in \eqref{eq:5Dvww},
which decomposes as
\begin{align}\label{eq:Gammaij}
64\, L_{ij} &=
	- \tfrac{1}{3} \veps_{ik} R(\cV)^{ab\,k}{}_j \, 
		\Big(\ri \bar X^0 T_{ab}{}^{mn} \veps_{mn}
		- \ri F_{ab}^{-}{}^0 + \HC \Big)\,|X^0|^{-2}
	\eol & \quad
	+ \tfrac{1}{12} Y_{ij}{}^0 \,\Big(\ri \,\bar X^0 T^{ab \,kl} \veps_{kl} F_{ab}^-{}^0
		- \ri \,(F_{ab}^-{}^0)^2 + \HC\Big)\, |X^0|^{-4} 
	\eol & \quad
	- \tfrac{2}{3} \ri \,\mathcal{D}^a \ln (X^0 / \bar X^0) \,\mathcal{D}_a (Y_{ij}{}^0 / |X^0|^2)~.
\end{align}
This expression includes all the bosonic contributions to $L_{ij}$.
Now let us calculate the same contribution from the $4D$ superspace action
\eqref{eq:4Dvww_super}. It helps to rewrite the action as
\begin{align}\label{eq:4Dproto}
\frac{\ri}{64} \int \rd^4x\, \rd^4\q\, \cE\, \frac{c_I X^I}{X^0} \Phi~, \qquad
\Phi = W^{\alpha\beta} W_{\alpha\beta} - \tfrac{1}{3} \mathbb{T}(\ln\bar X^0)
\end{align}
and express the component action in terms of the components of $\Phi$.
For example, the contribution to $L_{ij}$ is given by
\begin{align}\label{eq:4DprotoGij}
64 \,L_{ij} = \frac{\ri}{2} \frac{Y_{ij}{}^0}{(X^0)^2}\, A\vert_\Phi
	- \frac{\ri}{2} \frac{1}{X^0} B_{ij}\vert_\Phi + \HC
\end{align}
The components of $\Phi$ can then be calculated as
\begin{align}
  \label{eq:PhiAB}
  A\vert_\Phi &= \tfrac{1}{32} (T_{ab}{}^{ij} \veps_{ij})^2 -
  \tfrac{1}{3} A\vert_{\mathbb T(\ln\bar X^0)} 
	\eol
	&= \tfrac{1}{96} (T_{ab}{}^{ij} \veps_{ij})^2
	+ (\bar X^0)^{-1} \Big(
		\tfrac{2}{3} \Box_\mathrm{c} X^0
		+ \tfrac{1}{12} T^{ab\,ij} \veps_{ij} F^{-}_{ab}{}^0\Big)
	\eol & \quad
	+ (\bar X^0)^{-2} \Big(
		\tfrac{1}{6} (F^{+}_{ab}{}^0-\tfrac14X^0 T_{ab\,ij}\veps^{ij})^2
		- \tfrac{1}{12} (Y_{ij}{}^0)^2
		\Big)~, \eol
                B_{ij}\vert_\Phi &=\varepsilon_{ik}R({\cal V})_{ab}{}^k{}_{j}
		\Big\{\tfrac{1}{2} T^{ab\, kl} \veps_{kl}
		+ \tfrac{2}{3} (F^{+}_{ab}{}^0
		- \tfrac14 X^0 T_{ab\,kl}\veps^{kl}) \,(\bar X^0)^{-1}
                \Big\}
                \nonumber\\
                &\quad
                +\tfrac{2}{3}(\Box_\mathrm{c}+3D)\bigg(\frac{Y_{ij}{}^0}{\bar{X}^0}\bigg)~.
\end{align}
A straightforward calculation leads to $L_{ij}$ as in \eqref{eq:Gammaij}.
As already mentioned, this nearly guarantees equivalence of the
final expressions, but we will check some additional terms to
marshal further evidence.

Let us now analyze the second term $E_{\mu\nu}$ of the $4D$ Lagrangian 
\eqref{eq:4Dvww_gen}. We will check only a subset of contributions.
One obvious source is terms involving $\wt F_{AB}{}^I$ whose
decomposition in $4D$ tangent space indices
yields $F_{ab}{}^I$. These give contributions to the
$4D$ Lagrangian of the form
\begin{align}\label{eq:Fterms1}
&- \tfrac{1}{2} \,c_I \, F_{ab}{}^I \Big[
	 \tfrac{3}{16} \wt R(M)_{CD}{}^{ab}\, \wt T^{CD}
	+ \wt T^{ab} \Big(\wt D + \tfrac{3}{8} (\wt T_{CD})^2\Big)
	- \tfrac{9}{2} \wt T^{aC} \wt T_{CD} \wt T^{Db}\Big] |X^0|^{-1} 
	\eol & 
+ \tfrac{3}{8}\,\ri \,\veps^{abCDE} c_I F_{ab}{}^I 
	\Big(
		\wt T_{DF} \wt {\mathcal{D}}^F \wt T_{DE}
		+ \tfrac{3}{2} \wt T_{CF} \wt{\mathcal{D}}_D \wt T_E{}^F 
	\Big)\,|X^0|^{-1}~.
\end{align}
We will discuss how to simplify this expression shortly.
The other contributions come from the Chern-Simons term,
which gives
\begin{align}
-\tfrac{1}{64} \,\ri\,\veps^{abcd} \,c_I \,W_a{}^I \Big(
	\wt R(M)_{bc}{}^{EF} \wt R(M)_{d5\, EF}
	+ \frac{1}{3} \wt R(\cV)_{bc}{}_i{}^j \wt R(\cV)_{d5}{}_j{}^i
	\Big)\, |X^0|^{-1}~.
\end{align}
This can be rearranged to
\begin{align}\label{eq:Fterms2}
&-\tfrac{1}{64} \,\ri\,\veps^{abcd}  c_I F_{ab}{}^I
	\Big(
	\tfrac{1}{8} R_{cd}{}^{ef} F_{ef}{}^0 |X^0|^2
	+ \tfrac{1}{128} (F_{ef}{}^0)^2 F_{cd}{}^0
	+ \tfrac{1}{64} F^{ef\, 0} F_{ce}{}^0 F_{df}{}^0\Big) |X^0|^{-4}
\eol & \quad
+ \tfrac{1}{192} \,\ri\,\veps^{abcd}  c_I F_{ab}{}^I
	\Big(
	\tfrac{1}{4} \,\veps^{jk}\, R(\cV)_{cd}{}^{i}{}_k\, Y_{ij}{}^0\,|X^0|^2
	+ \tfrac{1}{32} F_{cd}{}^0 (Y_{ij}{}^0)^2
	\Big)\,|X^0|^{-4}
\end{align}
up to terms involving derivatives of $|X^0|$, which from now
on we will neglect to keep our expressions simpler.
It will be useful to neglect other terms in \eqref{eq:Fterms1}.
For example, expressions involving $\wt T_{a5}$ appear in nearly
every term, often in multiple ways (e.g. from the $5D$ spin connection),
so it will be convenient to set $\wt T_{a5}$ to zero, which
amounts to discarding $\mathcal{D}_a \ln (X^0 / \bar X^0)$.
We will also ignore all terms involving $F_{ab}{}^0$ that
also contain a factor of $T_{cd}{}^{ij}$, $T_{cd}{}_{ij}$ or another
$F_{cd}{}^0$. These conditions together allow us to focus on
only the first line of \eqref{eq:Fterms1}. Proceeding, we find that
the first line reduces to
\begin{align}\label{eq:Fterms1v2}
- \tfrac{1}{2} c_I \, F_{ab}{}^I \Big[
	 \tfrac{3}{16} \wt R(M)_{cd}{}^{ab}\, \wt T^{cd}
	+ \wt T^{ab} \Big(\wt D + \tfrac{3}{8} (\wt T_{cd})^2\Big)
	- \tfrac{9}{2} \wt T^{ac} \wt T_{cd} \wt T^{db}\Big]\, |X^0|^{-1}~.
\end{align}
Now we combine this with \eqref{eq:Fterms2} and find
the coefficient of $c_I F^{ab\, I}$ to be
\begin{align}\label{eq:Gab}
-64 \,\ri \,\tilde E_{ab}
	&\sim \tfrac{1}{2} \ri\, \cC_{abcd} \,T^{cd\, ij} \veps_{ij}\, (X^0)^{-1}
	+ \tfrac{1}{3} \ri\, \veps_{i k} \,R({\cal V})^{-}_{ab}{}^k{}_j \,Y^{ij\, 0}\, |X^0|^{-2}
	\eol & \quad
	+ \tfrac{4}{3} \ri\, (\cR_a{}^c -\tfrac{1}{4} \delta_a{}^c \cR) \,F_{cb}^+{}^0\, |X^0|^{-2}
	+ \tfrac{1}{9} \ri \,\cR (F_{ab}^-{}^0 + \tfrac{1}{2} \bar X^0 T_{ab}{}^{ij} \veps_{ij}) \,|X^0|^{-2}
	\eol & \quad
	- \tfrac{2}{3} \ri\, D \,(F_{ab}^-{}^0 - \bar X^0 T_{ab}{}^{ij} \veps_{ij}) \,|X^0|^{-2}
	- \tfrac{1}{12} \ri\, (Y_{ij}{}^0)^2 \, \big(F_{ab}^-{}^{0}
	- \tfrac{1}{2} \bar X^0 T_{ab}{}^{ij} \veps_{ij} \big)\, |X^0|^{-4}
	\eol & \quad
	- \tfrac{1}{192} \ri\, T_{ab}{}^{ij} \veps_{ij} \,(T_{cd}{}^{kl} \veps_{kl})^2\, \bar X^0 (X^0)^{-2}
	- \tfrac{1}{64} \ri\, T_{ab}{}^{ij} \veps_{ij}\, (T_{cd\, kl} \veps^{kl})^2\, (\bar X^0)^{-1}
	+ \HC
\end{align}
up to the terms we neglected. Keep in mind that $\tilde E_{ab}$ is imaginary
so the above expression is actually real.
To extract the corresponding terms from the $4D$ Lagrangian \eqref{eq:4Dvww_super},
we return to \eqref{eq:4Dproto}, where
\begin{align}\label{eq:4DprotoGab}
-64 \,\ri\, \tilde E_{ab}
	&= -\frac{\ri}{X^0} F^-_{ab}\vert_{\Phi}
	+ \frac{1}{(X^0)^2} \Big(
	\ri F^{-\,0}_{ab}
	- \tfrac{1}{4} \ri \,\bar{X}^0 T_{ab}{}^{ij} \veps_{ij}
	+ \tfrac{1}{4} \ri\, X^0 T_{ab\,ij} \veps^{ij}
	\Big) \,A\vert_{\Phi}
	+ \HC
\end{align}
The result for $A\vert_{\Phi}$ was given in \eqref{eq:PhiAB}.
The expression for $F_{ab}^-\vert_\Phi$ is
\begin{align}\label{eq:PhiFab}
F^-_{ab}\vert_\Phi
	&= -\tfrac{1}{2} {\cal R}(M)^{cd}{}_{ab} \,T_{cd}{}^{ij} \veps_{ij}
	-\tfrac{1}{3} \veps_{ij} T_{ab}{}^{ij} \,\Box_\mathrm{c}\ln\bar{X}^0
	+\tfrac{1}{3} R({\cal V})^{-}_{ab}{}^i{}_k Y^{jk\,0} \veps_{ij}\, (\bar{X}^0)^{-1}
	\eol & \quad	
	-\tfrac{1}{24} T_{ab}{}^{ij} T_{cd\, ij}(F^{cd\,+\,0}-\tfrac14 X^0T^{cd}{}_{kl} \veps^{kl}) (\bar{X}^0)^{-1}
	\eol & \quad	
	+ \tfrac{1}{3} (\delta_a{}^{[c} \delta_b{}^{d]} - \tfrac{1}{2} \veps_{ab}{}^{cd}) \Big[
		4 D_c D^e \Big(\frac{F^{+}_{ed}{}^0 -\tfrac14X^0T_{ab\,ij} \veps^{ij}}{\bar{X}^0}\Big)
		- D_c D^e T_{ed}{}^{ij} \veps_{ij}
		\eol & \qquad\qquad\qquad
		+ D^e\ln\bar{X}^0\,D_c T_{de}{}^{ij} \veps_{ij}
		+ D_c\ln\bar{X}^0 D^e T_{ed}{}^{ij} \veps_{ij}
	\Big]
\end{align}
A careful calculation, keeping only the terms discussed, reproduces \eqref{eq:Gab}.

Let us now analyze the last term $G$ of the $4D$ Lagrangian 
\eqref{eq:4Dvww_gen}.
Because of the complexity of the full expression, we
will only look at a small number of characteristic terms.
We begin with all terms involving the $4D$ SU(2) curvature tensor,
which arise only from the second and third lines of \eqref{eq:5Dvww}.
These are
\begin{align}\label{eq:Gterms1}
128 \, X^0 \,G &\sim
	- \tfrac{1}{3}\,\ri \,R(\cV)_{ab}^+{}^i{}_j R(\cV)^{ab+}{}^j{}_i
	- \ri R(\cV)_{ab}^-{}^i{}_j R(\cV)^{ab\,-}{}^j{}_i
	\eol & \qquad
	+ \tfrac{1}{8} R(\cV)_{ab}{}^j{}_k \veps^{ki} \,Y_{ij}{}^0
		\Big(\tfrac{4}{3}\,\ri \,\bar X^0\, T^{ab\,mn} \veps_{mn}
		+ \tfrac{8}{3} \,\ri F^{ab\,-\,0} + \HC\Big) |X^0|^{-2}
\end{align}
Next, we collect all terms involving the $4D$ auxiliary field $D$
that do not involve derivatives of $X^0$ or $\bar X^0$.
These arise only from $5D$ terms
involving $\wt D$ and are given by
\begin{align}\label{eq:Gterms2}
128\, X^0\, G &\sim
	-\tfrac{32}{3} \ri\, D^2
	+ \ri \,D \Big[
		\tfrac{1}{6} \frac{\bar X^0}{X^0} (T_{ab}{}^{ij} \veps_{ij})^2
		+ \tfrac{1}{6} \frac{X^0}{\bar X^0}(T_{ab\, ij}\veps^{ij})^2
		- \tfrac{2}{3} F_{ab}^-{}^0 T^{ab}{}^{ij} \veps_{ij} (X^0)^{-1}
	\eol & \qquad\qquad
		+ (F_{ab}^-{}^0)^2 |X^0|^{-2}
		+ \tfrac{1}{3} (F_{ab}^+{}^0)^2 |X^0|^{-2}
		+ \tfrac{8}{9} \cR
		- \tfrac{4}{3} (Y_{ij}{}^0)^2 |X^0|^{-2}
	\Big]~.
\end{align}
Finally, we include all expressions quadratic in
the $4D$ Riemann tensor as well as the terms $(Y_{ij}{}^0)^4$
and $\cR (Y_{ij}{}^0)^2$. These are easily deduced from the $5D$ Lagrangian
because they arise only from the second and third lines as well as
the term involving $\wt D^2$. The result is
\begin{align}\label{eq:Gterms3}
128\, X^0\,G &\sim
	-2 \ri\, \cC_{ab}^-{}^{cd} \cC_{cd}^-{}^{ab}
	- \tfrac{2}{3} \ri\, (\cR_{ab})^2
	+ \tfrac{4}{27} \ri\, \cR^2
	- \tfrac{1}{24} \ri (Y_{ij}{}^0)^4 |X^0|^{-4}
	+ \tfrac{1}{18} \ri\, \cR\, (Y_{ij}{}^0)^2 |X^0|^{-2}~.
\end{align}
These three sets of terms,  \eqref{eq:Gterms1}--\eqref{eq:Gterms3},
constitute a useful characteristic set. They can be found within
the $4D$ Lagrangian \eqref{eq:4Dproto}, for which $G$ is given by
\begin{align}\label{eq:4DG}
128 \,G &=
	-\frac{\ri}{X^0}\, C\vert_\Phi
	- \frac{\ri}{2 (X^0)^2} Y^{ij\,0} \,B_{ij}\vert_\Phi
	- \frac{\ri}{4\bar X^0}T^{ab}{}_{ij}\veps^{ij}\, F^+_{ab}\vert_\Phi
	+ \frac{\ri}{(X^0)^2}\big(
		F^{ab\,-\,0}
		- \tfrac14\bar X^0 T^{ab\,ij} \veps_{ij}
		\big)\, F^-_{ab}\vert_\Phi
	\eol & \quad
	- \frac{\ri}{(X^0)^2} \Big[
		2 \Box_\mathrm{c}\bar{X}^0
		+ \tfrac14 ( F^{+\,0}_{ab}-\tfrac14\,X^0T_{ab\, ij} \veps^{ij})T^{ab}{}_{kl} \veps^{kl}
		- \frac{1}{2\,X^0}Y_{ij}{}^0\,Y^{ij\,0}
		\eol & \qquad\qquad\qquad
		+\frac{1}{X^0}(F^{-\,0}_{ab} -\tfrac14\bar{X}^0\,T_{ab}{}^{ij} \veps_{ij})^2 \Big]\, A\vert_\Phi
	\eol & \quad
	- 2\ri \,\Box_\mathrm{c} \Big(\frac{\bar{A}\vert_\Phi}{\bar X^0}\Big)
	+ \frac{\ri}{4(\bar X^0)^2}\,T^{ab}{}_{ij} \veps^{ij} (F^{+0}_{ab}-\tfrac14X^0\,T_{ab\,kl}\veps^{kl}) \,\bar A\vert_\Phi~.
\end{align}
The expressions for all of the bosonic components of $\Phi$ have
been given except for $C\vert_\Phi$. It is rather lengthy,
so we refer to \cite{Butter:2013lta} where it was evaluated in detail.

\section*{Acknowledgements}
This work is supported in part by the ERC Advanced
Grant no. 246974, {\it ``Supersymmetry: a window to non-perturbative physics''}.

\providecommand{\href}[2]{#2}

\end{document}